\documentclass{SciPost}

\usepackage[utf8]{inputenc}
\usepackage{graphicx}
\usepackage{xcolor}
\usepackage{amsmath}
\usepackage{multirow}
\usepackage{array}
\usepackage{hyperref}
\usepackage{comment}
\usepackage{float}
\usepackage{amsmath}
\usepackage{tabularx,graphicx}
\usepackage{epstopdf}
\usepackage{latexsym}
\usepackage{color, colortbl}
\usepackage{psfrag}
\usepackage{bbm}
\usepackage{bm}
\usepackage{titlesec}
\usepackage{dsfont}
\usepackage{feynmp}
\usepackage{slashed}
\usepackage{subcaption}
\usepackage{enumitem}
\usepackage{multirow}
\usepackage{subcaption}
\usepackage[newcommands]{ragged2e}
\usepackage[normalem]{ulem}
\renewcommand{\vec}[1]{{\boldsymbol{#1}}}
\usepackage{braket}
\usepackage{subcaption}
\usepackage{cases}
\usepackage{slashbox}
\usepackage{cellspace} 
\def \q {{\vec q}}

\def \beq {\begin{eqnarray}}
\def \eeq {\end{eqnarray}}

\def \vp{\varphi}

\def \nn {\nonumber}

\newcommand{\change}[1]{#1}
\newcommand{\rd}{{\rm d}}
\newcommand{\sgn}{{\rm sgn\,}}

\newcommand{\vn}[1]{{\left|\vec{#1}\right|}}
\newcommand{\calN}{{\mathcal N}}

\newcommand{\calF}{{\mathcal F}}

\newcommand{\calL}{{\mathcal L}}

\newcommand{\calO}{{\mathcal O}}
\renewcommand{\Re}{\text{Re}}
\renewcommand{\Im}{\text{Im}}

\usepackage{tikz-feynman}
\usetikzlibrary{calc}
\usetikzlibrary{arrows}
\tikzset{
  mid arrow/.style={postaction={decorate,decoration={
        markings,
        mark=at position .575 with {\arrow[#1]{stealth}}
      }}},
  near arrow/.style={postaction={decorate,decoration={
        markings,
        mark=at position .275 with {\arrow[#1]{stealth}}
      }}},
   far arrow/.style={postaction={decorate,decoration={
        markings,
        mark=at position .800 with {\arrow[#1]{stealth}}
      }}},
   boson/.style={decorate, draw=black,
    decoration={snake,amplitude=1pt, segment length=5pt},
      },
      phonon/.style={decorate, draw=red,
    decoration={snake,amplitude=1pt, segment length=5pt},
      },
   mid triangle/.style={postaction={decorate,decoration={
        markings,
        mark=at position .575 with {\arrow[#1]{triangle 45}}
      }}}
}

\usepackage{tikz-feynman}
\usetikzlibrary{arrows}
\usetikzlibrary{arrows.meta, positioning,shapes.geometric}
\tikzset{
  mid arrow/.style={postaction={decorate,decoration={
        markings,
        mark=at position .575 with {\arrow[#1]{stealth}}
      }}},
  near arrow/.style={postaction={decorate,decoration={
        markings,
        mark=at position .275 with {\arrow[#1]{stealth}}
      }}},
   far arrow/.style={postaction={decorate,decoration={
        markings,
        mark=at position .800 with {\arrow[#1]{stealth}}
      }}},
   boson/.style={decorate, draw=black,
    decoration={snake,amplitude=1pt, segment length=5pt},
      },
   mid triangle/.style={postaction={decorate,decoration={
        markings,
        mark=at position .575 with {\arrow[#1]{triangle 45}}
      }}}
}
\tikzset{
    position label/.style={
       below = 3pt,
       text height = 1.5ex,
       text depth = 1ex
    },
   brace/.style={
     decoration={brace,raise=7pt},
     decorate
   }
}

\newmuskip\pFqmuskip

\newcommand*\pFq[6][8]{%
  \begingroup 
  \pFqmuskip=#1mu\relax
  \mathchardef\normalcomma=\mathcode`,
  \mathcode`\,=\string"8000
  \begingroup\lccode`\~=`\,
  \lowercase{\endgroup\let~}\pFqcomma
  {}_{#2}F_{#3}{\left[\genfrac..{0pt}{}{#4}{#5};#6\right]}%
  \endgroup
}
\newcommand{\pFqcomma}{{\normalcomma}\mskip\pFqmuskip}

\hypersetup{
    colorlinks,
    linkcolor={red!50!black},
    citecolor={blue!50!black},
    urlcolor={blue!80!black}
}

\usepackage[bitstream-charter]{mathdesign}
\DeclareSymbolFont{usualmathcal}{OMS}{cmsy}{m}{n}
\DeclareSymbolFontAlphabet{\mathcal}{usualmathcal}

\fancypagestyle{SPstyle}{
\fancyhf{}
\lhead{\colorbox{scipostblue}{\bf \color{white} ~SciPost Physics }}
\rhead{{\bf \color{scipostdeepblue} ~Submission }}

\fancyfoot[C]{\textbf{\thepage}}
}
\begin{document}

\pagestyle{SPstyle}

\begin{center}{\Large \textbf{\color{scipostdeepblue}{
Can electronic quantum criticality drive phonon-induced \\ linear-in-temperature resistivity?\\
}}}\end{center}

\begin{center}\textbf{
Haoyu Guo\textsuperscript{1$\star$} and
Debanjan Chowdhury\textsuperscript{1$\dagger$}
}\end{center}

\begin{center}
{\bf 1} Department of Physics, Cornell University, Ithaca, New York 14853, USA.
\\[\baselineskip]
$\star$ \href{mailto:haoyuguo@cornell.edu}{\small haoyuguo@cornell.edu}\,,\quad
$\dagger$ \href{mailto:debanjanchowdhury@cornell.edu}{\small debanjanchowdhury@cornell.edu}
\end{center}

\section*{\color{scipostdeepblue}{Abstract}}
\textbf{\boldmath{
Optical phonons naturally generate linear-in-temperature ($T$) resistivity in the equipartition regime, but their finite gap prevents this mechanism from surviving to asymptotically low temperatures. Here we analyze whether proximity to an electronic quantum critical point can remove this obstruction by strongly softening an optical phonon. We first derive a model-independent criterion for such softened phonons to control low-temperature transport. In addition to reducing the renormalized optical gap, the Landau-damped phonon must acquire a dynamical exponent $z_p>d$, where $d$ is the spatial dimension of the phonon, so that a sufficiently large thermally occupied phase space survives as $T\to 0$. We then analyze a concrete mechanism in which the phonon couples nonlinearly to long-wavelength electronic collective modes near a quantum critical point associated with an order parameter carrying zero center-of-mass momentum, and apply it to the Ising-nematic problem. Within a large-$N$ field theoretic formulation, the phonon softening is enhanced near criticality, but in the clean theory the resulting dynamics lies at or near the  boundary for asymptotic $T$-linear scattering. Including feedback from the softened phonon back onto the electronic critical sector further weakens the tendency toward robust low-temperature $T$-linear transport. Our results sharpen both the promise and the limitations of phonon-based explanations of strange-metal transport near electronic criticality.
}}

\vspace{\baselineskip}

\vspace{10pt}
\noindent\rule{\textwidth}{1pt}
\tableofcontents
\noindent\rule{\textwidth}{1pt}
\vspace{10pt}

\section{Introduction}
The electrical transport properties of strongly correlated metals remain among the most enigmatic phenomena in quantum materials research \cite{varmaRMP,SAHartnoll2022,DChowdhury2022a}. Particularly striking is the behavior of strange metals, whose dc electrical resistivity scales linearly with temperature ($T$) over an unusually broad window and often persists down to remarkably low energy scales before the onset of superconductivity in chemically diverse material families \cite{JANBruin2013}. This $T$-linear resistivity is frequently discussed in terms of an apparently universal, scale-invariant ``Planckian'' scattering rate set only by temperature and fundamental constants \cite{JZaanen2004}, suggesting possible links to quantum criticality \cite{SSachdev1999b} and, more broadly, to strongly interacting metallic states without long-lived quasiparticles \cite{hartnoll2018holographic,PWPrev}. The apparent connection to criticality is reinforced by the empirical observation that strange-metal transport often appears near quantum phase transitions associated with electronic ordering tendencies \cite{LTaillefer2010,matsuda,Keimer15}, including spontaneous rotational \cite{nematicreview} and translational symmetry breaking \cite{CProust2019}. At the same time, a scale-invariant local equilibration timescale does not by itself guarantee $T$-linear resistivity, which is ultimately controlled by momentum relaxation rather than local relaxation alone.

In crystalline solids, electron-phonon interactions provide the most familiar route to $T$-linear resistivity. Once the relevant phonon modes enter the equipartition regime, they act as an approximately quasi-elastic source of scattering \cite{ziman2001electrons}. This mechanism has even been argued to be relevant for parent states of high-temperature superconductors \cite{EHHwang2019,EJHeller2025}. However, several observations indicate that conventional phonons cannot by themselves account for the generic strange-metal phenomenology of correlated materials, including optimally doped cuprates \cite{CHMousatov2021}. First, the experimentally observed $T$-linear regime often extends to temperatures well below the relevant Debye or Bloch-Gr\"uneisen scales \cite{martin,GGrissonnanche2021,JAyres2021d,PGiraldo-Gallo2018,YCao2020,AJaoui2022}. Second, the repeated appearance of strange metallicity near electronic symmetry-breaking transitions is unlikely to be accidental, and already points to an essential role for low-energy electronic collective fluctuations \cite{LTaillefer2010,matsuda,IMHayes2016}. Third, correlated materials typically possess multiple optical and acoustic phonon branches with widely separated characteristic energies, making it unclear which modes, if any, can remain effective in the asymptotically low-$T$ regime \cite{SDasSarma2024}. These difficulties have motivated an extensive search for purely electronic mechanisms of $T$-linear transport, often involving special forms of disorder \cite{AAPatel2023,CLi2024,EEAldape2022,AAPatel2024a,AAPatel2024}, particular order-parameter dynamics \cite{DVElse2021,ZDShi2023}, glassy order \cite{NBashan2024,NBashan2025}, or umklapp-dominated relaxation channels \cite{XWang2019,PALee2021,PALee2024}.

Remarkably, there is substantial experimental evidence that optical phonons are themselves strongly renormalized in precisely the parameter regimes where electronic correlations are most singular. In cuprates, pnictides, and organic superconductors, Raman spectroscopy \cite{lsco_raman, lsco_raman2,liarokapisLatticeEffectsLa_rm2008, blumenroederPhononRamanScattering1987, Fe_11_raman, Fe_11_raman2} and inelastic neutron and X-ray scattering \cite{lsco_ixs1, birgeneauSoftphononBehaviorTransport1987, bradenElasticInelasticNeutrona, kimuraStructuralInstabilityAssociated2000, parkEffectsChargeInhomogeneities2011, parkEvidenceChargeCollective2014, wakimotoNeutronScatteringStudy2004} have reported pronounced softening of selected optical modes as a function of doping, pressure, or composition. These anomalies are typically strongest near structural, magnetic, or nematic instabilities, and are widely interpreted as signatures of strong coupling between lattice degrees of freedom and low-energy electronic order-parameter fluctuations. Strikingly, the same regions of the phase diagram often host the most pronounced strange-metallic transport \cite{kasaharaEvolutionNonFermiFermiliquid2010, nakaiUnconventionalSuperconductivityAntiferromagnetic2010, andoElectronicPhaseDiagram2004, legrosUniversalTlinearResistivity2019a}. While these observations do not by themselves show that softened phonons are responsible for the measured $T$-linear resistivity, they make it natural to revisit the role of phonons in a setting where their dispersion and characteristic energy scales are themselves shaped by electronic criticality. They also raise a complementary conceptual question for purely electronic theories: when experiments point to strong electron-phonon coupling \cite{EPCBSCCO,JZhang2017a,JZhang2018,CHMousatov2021}, under what conditions can phonons be consistently neglected in the first place?

\begin{figure}[htb]
    \centering
    \includegraphics[width=0.9\linewidth]{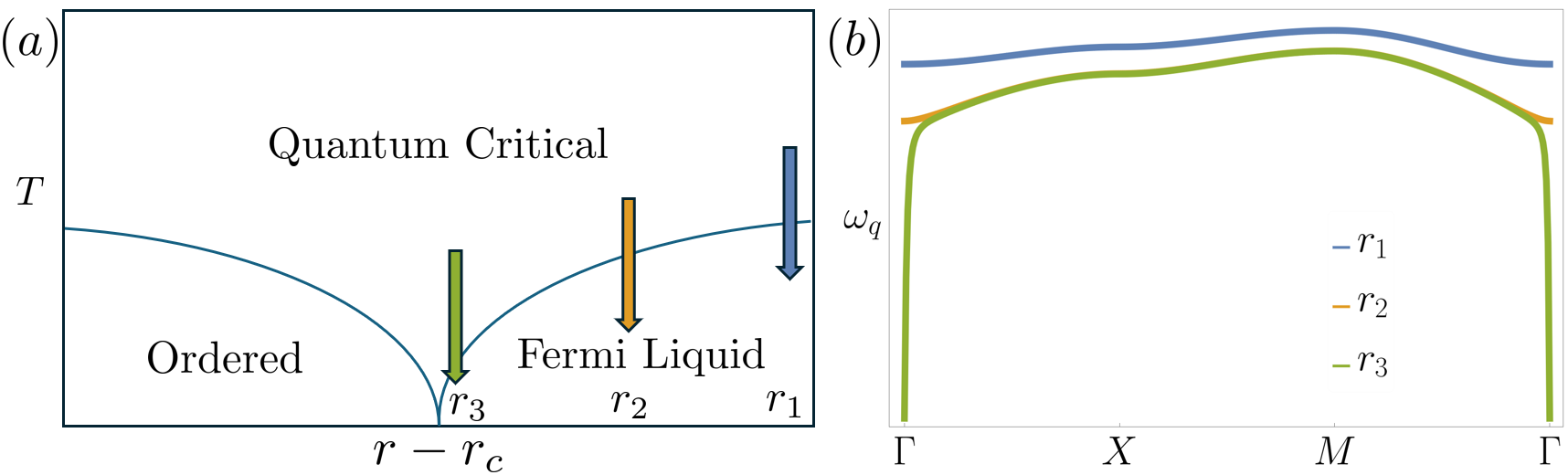}
    \includegraphics[width=0.6\linewidth]{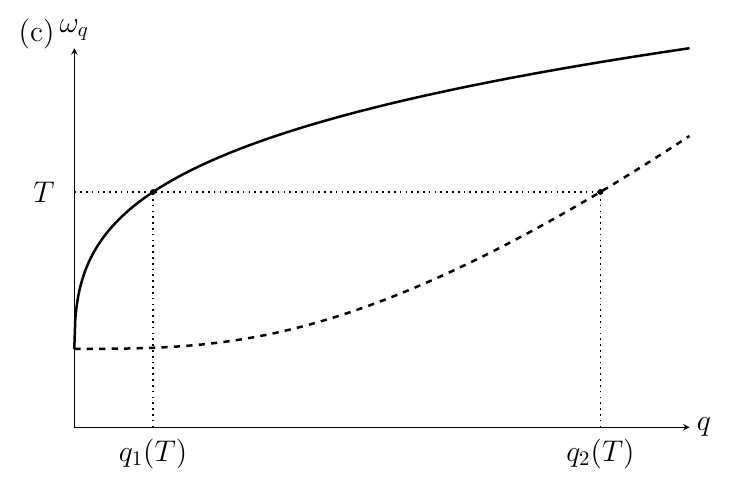}
    \caption{Schematic consequences of phonon softening near an electronic quantum critical point. (a) As the system is tuned toward criticality ($r_1\rightarrow r_2\rightarrow r_3$), an optical phonon mode softens and its long-wavelength dispersion is renormalized, increasing the density of low-energy phonon states near the $\Gamma-$point, as shown in panel (b). (c) For a renormalized phonon dispersion $\omega_{\vec q}\sim q^{z_p}$, the largest thermally occupied momentum scales as $q(T)\sim T^{1/z_p}$, so the phase-space volume of thermally excited phonons scales as $[q(T)]^d\sim T^{d/z_p}$, where $d$ is the spatial dimension and $z_p$ is the phonon dynamical exponent. A larger $z_p$ therefore produces a larger low-energy phonon phase space. The solid and dashed curves illustrate the qualitatively different cases $z_p<d$ and $z_p>d$, where the characteristic thermal momentum  scales are labelled by $q_1(T)$ and $q_2(T)$, respectively.  In the low-temperature limit, $q_2(T)\gg q_1(T)$ in the scaling sense, implying that there is more phonon phase-space participating in scattering in the $z_p>d$ case,  compared to the $z_p<d$ case.
}
    \label{fig:sketch}
\end{figure}

These observations motivate the central question of this paper, namely whether proximity to an electronic quantum critical point can soften an optical phonon enough to make phonon-mediated $T$-linear transport survive parametrically to low temperatures? To address this question, we study a theory of intertwined phonons and bosonic collective modes of electronic origin near symmetry-breaking quantum critical points. We focus on low-energy symmetry-allowed couplings between electrons, critical collective modes, and optical phonons, leaving the corresponding problem for acoustic phonons to future work. Our strategy is to separate the problem into two logically distinct parts. First, we ask in a model-independent way what kind of renormalized phonon dispersion is required for softened optical phonons to generate linear-in-$T$ single-particle and transport scattering rates. This leads to a general criterion involving the phonon dynamical exponent $z_p$ and spatial dimension $d$. Second, we analyze whether such a dispersion can arise microscopically from a concrete coupling between the phonon and an electronic quantum critical mode. In the examples studied here, the phonon softening is concentrated near the $\Gamma$ point (see Fig.~\ref{fig:sketch}), so converting the enhanced low-energy phonon phase space into resistivity additionally requires an umklapp-enabled transport mechanism. We therefore treat the quantum critical renormalization of the phonon spectrum and the transport consequences of that renormalized spectrum as two separate but connected steps.

The remainder of this manuscript is organized as follows. In Sec.~\ref{sec:electronphonon}, we formulate general criteria for when softened optical phonons can give rise to linear-in-$T$ single-particle and transport scattering rates. In particular, in the regime $T\gg \omega_D$ (renormalized optical phonon gap), we identify a necessary condition involving the phonon dynamical exponent $z_p$ and spatial dimension $d$, and we clarify the role of umklapp scattering in converting single-particle decay into resistivity. In Sec.~\ref{sec:softening}, we then analyze these criteria in a concrete model of phonon softening near a $\vec Q=0$ electronic quantum critical point, where the phonon couples nonlinearly to the critical electronic mode. Within a leading order in large $N$ field-theoretic formulation of the problem (e.g. using the Yukawa-SYK construction), we compute the renormalized phonon dispersion and show that this mechanism places the system at the threshold of linear-in-$T$ scattering. We end with a brief conclusion in Sec.~\ref{sec:conclusion} and leave all the technical details to a number of appendices.

\section{From renormalized phonon dispersion to resistivity}\label{sec:electronphonon}

This section isolates the transport problem from the microscopic origin of the phonon softening. We assume that some mechanism has already produced a renormalized optical phonon dispersion and ask a more general question: under what conditions do those softened phonons generate a low-temperature linear-in-$T$ electron scattering rate and, after momentum relaxation is included, a linear-in-$T$ resistivity? The answer depends not only on how small the renormalized optical gap becomes, but also on how the thermally occupied phonon phase space scales with temperature once Landau damping is included. For that reason, we analyze the transport problem first and defer the microscopic origin of the softening to Sec.~\ref{sec:softening}.

\change{The rest of this section is as follows. We first state the assumptions under which a renormalized optical phonon spectrum can be treated as an effective low-energy input. We then compute the electron-phonon single-particle scattering rate for several representative geometries, using the result to identify the central criterion for a parametrically broad low-temperature linear-in-$T$ regime: after Landau damping is included, the effective phonon dynamical exponent must satisfy $z_p>d$, where $d$ is the spatial dimensionality of the phonon dispersion. Finally, we explain the additional transport requirements, especially momentum relaxation and umklapp kinematics, that are needed for the same scaling to appear in the dc resistivity. }

\subsection{Preliminaries: Assumptions and Model}

Our goal in this section is to derive the transport consequences of a given renormalized phonon spectrum while keeping the assumptions as explicit as possible. We make three simplifying assumptions.
\begin{enumerate}
    \item For dc transport, the dominant electron-phonon interaction is the conventional deformation-potential coupling \cite{ziman2001electrons}, schematically $c^\dagger c X$, and it can be analyzed within standard Migdal-Eliashberg theory.

   \item In addition to the deformation-potential coupling, the phonon may couple to other low-energy electronic collective modes \cite{Guo2025}, as discussed in Sec.~\ref{sec:softening}. In this section, we do not analyze those couplings microscopically. Instead, we assume that their leading effect is already captured by a renormalized low-energy phonon dispersion, which we treat as an input. Any additional damping channels associated with these couplings are assumed not to parametrically dominate the electron-phonon Landau damping processes considered below.

 \item The transport lifetime can track the same parametric temperature dependence as the single-particle lifetime provided that the softened phonons can effectively participate in momentum relaxation. This requires, first, that the phonon system itself can exchange momentum efficiently with other degrees of freedom, so that phonon drag is not the dominant effect \cite{YGGurevich1989}; and second, that the relevant electron-phonon scattering processes are not parametrically suppressed by small-angle kinematics, for example because phonon-mediated umklapp scattering is operative \cite{PALee2021}. We will return to these conditions in Sec.~\ref{sec:umklapp}.
\end{enumerate}
With these assumptions, we consider electrons coupled to an already renormalized phonon mode, with a Lagrangian density $\calL=\calL_e+\calL_p+\calL_{ep}$,
\begin{subequations}\label{eq:action1}
  \begin{eqnarray}
   \calL_e & =& \int \rd\tau \left[\sum_{\vec{k},\sigma} c_{\vec{k},\sigma}^\dagger \left(\partial_\tau+\varepsilon_{\vec{k}}-\mu\right)c_{\vec{k},\sigma} + \sum_\sigma \int_\vec{x} V_\vec{x} c_{\vec{x},\sigma}^\dagger c_{\vec{x},\sigma} \right] +\dots \label{eq:Le1} \\
   \calL_{p} &=& \int\rd \tau \sum_a\sum_{\q}X_{a,-\vec{q}} \left(\partial_{\tau}^2+\omega_\vec{q}^2\right) X_{a,\vec{q}}\,,\label{eq:Lp1}\\
   \calL_{ep} &=& \lambda\int\rd \tau \sum_{\vec{k},\vec{q},\sigma,a} h^a_{\vec{k},\vec{q}}c^\dagger_{\vec{k+q/2},\sigma}c^{\phantom\dagger}_{\vec{k-q/2},\sigma} X_{a,\vec{q}}+\dots\label{eq:Lep1}
\end{eqnarray}
\end{subequations}
Here $c_\vec{k,\sigma}^\dagger$ creates an electron with momentum $\vec{k}$, spin $\sigma=\uparrow,\downarrow$, and energy $\varepsilon_\vec{k}$ relative to chemical potential $\mu$; we will use $k_F$ to denote the Fermi momentum and $v_F$ to denote the Fermi velocity; we use $\sum_{\vec{k}}=\int \rd^3\vec{k}/(2\pi)^3$ interchangeably. The disorder potential $V_\vec{x}$ satisfies $\overline{V_\vec{x}}=0$ and $\overline{V_\vec{x}V_\vec{x'}}=V^2 \delta(\vec{x}-\vec{x'})$. We mostly focus on the clean limit $V\to 0$, using weak disorder only as a source of residual resistivity and as a mechanism that suppresses phonon drag and relaxes hot-spot bottlenecks in umklapp scattering \cite{ARosch1999,PALee2021}. The field $X_a$ denotes the phonon displacement, where $a$ labels phonon branches. The electron-phonon coupling channel that we focus here is of the Yukawa coupling form as defined in Eq.\eqref{eq:Lep1}, where $\lambda$ is the coupling constant and $h_\vec{k,q}^a$ is the form factor. 

The $\dots$ in Eq.~\eqref{eq:Le1} represent electron-electron interactions, which may be decoupled into a bosonic collective mode $\varphi$ of electronic origin. Near an electronic quantum critical point, that mode can become soft and, through the additional couplings contained in the $\dots$ of Eq.~\eqref{eq:Lep1}, renormalize the long-wavelength optical phonon. The microscopic origin of this renormalization will be analyzed in Sec.~\ref{sec:softening}. In the present section, however, we do not commit to a specific mechanism. Instead, we parameterize its net low-energy effect through a renormalized phonon dispersion and ask how the resulting softened phonons contribute to electronic lifetimes and transport. This leads us to three representative  scenarios, distinguished only by the dimensionality of the electronic and phonon sectors and by the corresponding low-energy form of $\omega_{\vec q}$:
\begin{itemize}
\item \textbf{Scenario (i)}   both the electronic and phonon sectors are quasi-two-dimensional, with $\omega_\vec{q}^2=\omega_D^2+C^2\vn{q_\text{2D}}^{z_p-1}$;
\item \textbf{Scenario (ii)}  the electronic sector is quasi-two-dimensional while the phonon disperses in three dimensions, with $\omega_\vec{q}^2=\omega_D^2+C^2\vn{q_\text{2D}}^{z_p-1}+c_z^2q_z^2$;
\item \textbf{Scenario (iii)}  both sectors disperse in three dimensions, with $\omega_\vec{q}^2=\omega_D^2+C^2\vn{q}^{z_p-1}$.
\end{itemize}
Here $z_p$ is the effective phonon dynamical exponent that emerges after softening, and $\omega_D$ is the renormalized phonon gap.

For the explicit estimates below, we specialize to a cylindrical Fermi surface in the quasi-two-dimensional cases and to a spherical Fermi surface in the three-dimensional case. The corresponding density of states per spin is $\calN_\text{2D}=k_F/(2\pi v_F a_z)$ for the quasi-2D geometry and $\calN_\text{3D}=k_F^2/(2\pi^2 v_F)$ in 3D. We further simplify the analysis by focusing on a single phonon branch that is strongly renormalized, suppressing the branch index $a$ when no confusion arises, and by taking the deformation-potential form factor to be smooth and nonvanishing on the relevant parts of the Fermi surface, so that for scaling purposes we may set $h_{\vec{k},\vec{q}}^a\to 1$. With these geometric and kinematic simplifications in place, the rest of this section determines how the three possible low-energy realizations of $\omega_{\vec q}$ translate into electron lifetimes and transport.

\subsection{Electron-phonon scattering rate}

We begin by recalling the standard picture of electron-phonon scattering in a conventional metal. Consider an electronic Fermi surface coupled to a weakly dispersing optical phonon with frequency $\omega_D$. The electron-phonon scattering rate, $\Gamma_\text{ep}$, then exhibits two familiar regimes. When $T\ll \omega_D$, the system lies in an activated regime in which both the phonon occupation and $\Gamma_\text{ep}$ are exponentially suppressed, $\propto e^{-\omega_D/T}$. When $T\gg \omega_D$, the system enters the equipartition regime, in which the phonon occupation and $\Gamma_\text{ep}$ both scale as $\propto T/\omega_D$. However, this simple picture is modified once the phonon becomes sufficiently soft that its damping by particle-hole excitations can no longer be neglected. The resulting Landau damping alters the standard optical-phonon crossover in three important ways:
\begin{enumerate}[label=(\alph*)]
    \item The activated low-temperature behavior of $\Gamma_\text{ep}$ is replaced by a power law, $\Gamma_\text{ep}\propto T^2/\omega_D^4$. The $T^2$ dependence is the familiar Fermi-liquid result, which here arises because an off-shell phonon mediates an effective interaction between electrons.

    \item Landau damping enlarges the phase space of phonons that behave as effectively classical scatterers. For a phonon with frequency $\omega_{\vec q}$ and momentum $\vec q$, the usual equipartition criterion $T\gg \omega_{\vec q}$ is relaxed to $T\gg v_F |\vec q|\,\omega_{\vec q}^2/\gamma$, where $\gamma$ is the Landau-damping coefficient defined below in Eq.~\eqref{eq:Delta_Pi_X}. In the low-energy limit, this condition is parametrically easier to satisfy.

    \item Because the damping-assisted equipartition regime depends on both the frequency and momentum of the phonon, the set of thermally active phonons becomes sensitive to the detailed form of the renormalized dispersion. As a result, the single crossover at $T\sim \omega_D$ is replaced by a richer structure with multiple temperature-dependent crossovers, whose form depends on the low-energy phonon dynamics.
\end{enumerate}

We now turn this qualitative picture into a scaling analysis of the electron self-energy in the theory introduced in Eqs.~\eqref{eq:Le1}-\eqref{eq:Lep1}. We work within the Migdal-Eliashberg approximation, appropriate when the phonons are parametrically slower than the electrons so that vertex corrections to the electron-phonon interaction can be neglected. In the large-$N$ Yukawa-SYK framework reviewed in Appendix~\ref{sec:YSYK}, this approximation can be made formally controlled. The full electron self-energy is $\Sigma=\Sigma_\text{ee}+\Sigma_\text{ep}$, where $\Sigma_\text{ee}$ and $\Sigma_\text{ep}$ arise from electron-electron and electron-phonon interactions, respectively. For the moment we continue to neglect impurity scattering. Within the Migdal-Eliashberg approximation, the electron-phonon contribution can be written as
\begin{equation}\label{eq:Sigma_ep1}
  \Sigma_\text{ep}(i\omega,\vec{k})=\lambda^2 T\sum_{\Omega} \int \frac{\rd^3\vec{q}}{(2\pi)^3} G(i\omega+i\Omega,\vec{k}+\vec{q})D_X(i\Omega,\vec{q})\,.
\end{equation}
Here $G(i\omega,\vec{k})^{-1}=i\omega-\xi_\vec{k}-\Sigma(i\omega)$ is the electron Green's function, with $\xi_\vec{k}=\varepsilon_\vec{k}-\mu$, and $D_X(i\Omega,\vec{q})^{-1}=\Omega^2+\omega_\vec{q}^2-\Pi_X(i\Omega,\vec{q})$ is the phonon Green's function. As long as $\Sigma(i\omega,\vec{k})$ depends only weakly on $\xi_\vec{k}$ near the Fermi surface, the resulting scaling of $\Sigma_\text{ep}$ is unchanged. We verify this explicitly in Appendix~\ref{sec:funcI}, where the relative dependence is shown to be suppressed by $|\xi_\vec{k}|/(k_F v_F)\ll1$.

The leading contribution to the phonon self-energy, $\Pi_X$, comes from Landau damping due to excitations near the Fermi surface,
\begin{equation}\label{}
  \Pi_X(i\Omega,\vec{q})=-\lambda^2 T\sum_{\omega}\int\frac{\rd^3\vec{k}}{(2\pi)^3} G(i\omega+i\Omega,\vec{k}+\vec{q}) G(i\omega,\vec{k})\,.
\end{equation}
It is convenient to separate this as $\Pi_X(i\Omega,\vec{q})=\Pi_X(i\Omega=0,\vec{q})+\Delta \Pi_X(i\Omega,\vec{q})$. The static piece is absorbed into the renormalized phonon dispersion introduced in Sec.~\ref{sec:model_softening}, while the remaining dynamical piece describes dissipation of phonons, which comes from Landau damping:
\begin{equation}\label{eq:Delta_Pi_X}
  \Delta \Pi_X(i\Omega,\vec{q})=-\gamma\frac{|\Omega|}{\eta_\vec{q}}\,,
\end{equation}
where the explicit forms of $\gamma$ and $\eta_\vec{q}$ depend on the Fermi-surface geometry and will be given below.

To quantify the electron-phonon scattering rate, we analytically continue $\Sigma_\text{ep}$ to real frequency, $i\omega\to \omega+i0$, and take its imaginary part:
\begin{equation}\label{eq:Gammaep1}
\begin{split}
 & \Gamma_\text{ep}=-2\Im \Sigma^R_\text{ep}(\omega=0,\vn{k}=k_F)=\\
  &-\lambda^2\int\frac{\rd^3 \vec{q}}{(2\pi)^3} \int \frac{\rd z}{2\pi} \frac{1}{\sinh \beta z} A_F(z,\vec{k}+\vec{q})A_X(z,\vec{q})\,.
\end{split}
\end{equation}
Here $A_F$ and $A_X$ are the electron and phonon spectral functions, respectively. Using Eq.~\eqref{eq:Delta_Pi_X}, we obtain
\begin{equation}\label{}
  -A_X(z,\vec{q})=\frac{2z \gamma \eta_\vec{q}}{z^2 \gamma^2+\eta_\vec{q}^2(z^2-\omega_\vec{q}^2)^2}\,.
\end{equation}

\begin{figure*}[htb]
  \centering
  \begin{subfigure}[t]{0.495\textwidth}
  \includegraphics[width=\textwidth]{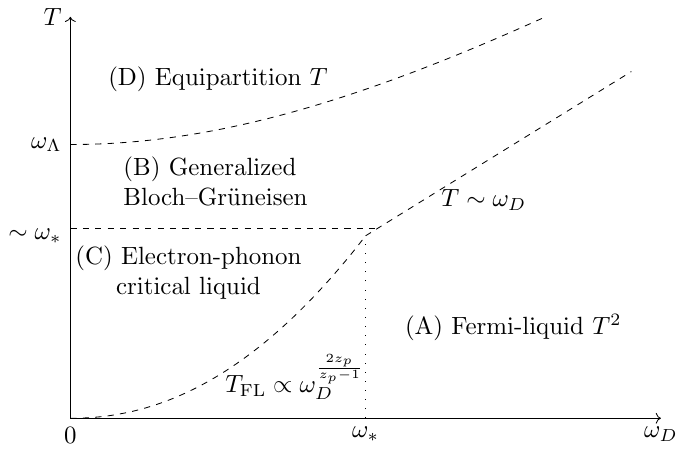}
  \caption{$1\leq z_p\leq z_c$}
  \end{subfigure}
  \begin{subfigure}[t]{0.495\textwidth}
  \includegraphics[width=\textwidth]{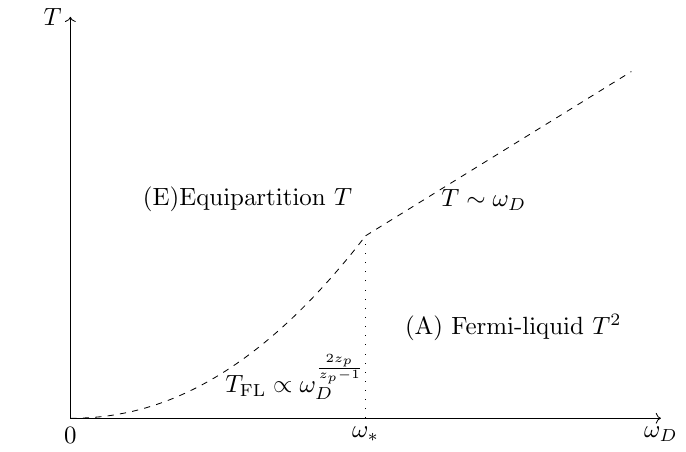}
  \caption{$z_p>z_c$}
  \end{subfigure}
  
  \caption{Crossover structure of the electron-phonon scattering rate $\Gamma_\text{ep}$ as a function of temperature $T$ and renormalized optical-phonon gap $\omega_D$. The qualitative behavior depends on whether the effective phonon dynamical exponent $z_p$ is smaller or larger than the critical value $z_c=d$, where $d$ is the spatial dimensionality of the phonon dispersion ($d=2$ or $3$ in the three scenarios discussed in the text). (a) For $1\le z_p\le z_c$, the scattering rate exhibits four distinct regimes, labeled (A,B,C,D). (b) For $z_p>z_c$, the regimes (B,C,D) merge into a single broadened equipartition-like regime (E), in which $\Gamma_\text{ep}$ becomes linear in $T$. The crossover scales $\omega_*$ and $\omega_\Lambda$ are defined in Eqs.~\eqref{eq:omega_Lambda} and \eqref{eq:omega_star}, respectively.}
  \label{fig:selfenergy1}
\end{figure*}

The formal expressions above apply to any Fermi-surface geometry and any renormalized phonon dispersion, but their scaling consequences depend on the dimensionality of the electron and phonon sectors. In the three scenarios introduced in Sec.~\ref{sec:model_softening}, this leads to the crossover structures summarized in Fig.~\ref{fig:selfenergy1} and Table~\ref{tab:selfenergy1}. The key distinction is whether the renormalized phonon dynamical exponent satisfies $z_p\le z_c$ or $z_p>z_c$, where $z_c=d$ is set by the spatial dimensionality of the phonon dispersion. For $z_p\le z_c$, the scattering rate passes through four regimes, labeled (A,B,C,D) in Fig.~\ref{fig:selfenergy1}(a). For $z_p>z_c$, by contrast, the regimes (B,C,D) merge into a single broadened equipartition-like regime (E), shown in Fig.~\ref{fig:selfenergy1}(b), in which $\Gamma_\text{ep}$ becomes linear in $T$. In the next three subsections, we analyze these cases one by one, deferring the detailed derivations to Appendix~\ref{sec:funcI}.

\change{Before presenting the case-by-case formulas, let us also discuss how the list of regimes should be interpreted. Regime (A) is the low-temperature Fermi-liquid regime controlled by virtual, effectively gapped phonons. Regimes (B) and (C) describe thermally occupied phonons before the full phonon band has entered equipartition, with the distinction between them set by whether the relevant modes are underdamped or overdamped. Regime (D) is the conventional high-temperature equipartition regime. Regime (E), which appears only when $z_p>d$, is the special softened-phonon regime in which the lowest-energy phonons already provide enough phase space to give linear-in-$T$ scattering without waiting for the entire phonon bandwidth to be thermally populated.}

\begin{table}[htb]
  \centering
  \begin{tabular}{|c|Sc|Sc|Sc|Sc|Sc|}
    \hline
    \backslashbox{Scenario}{Regime} & A & B & C & $D$ & $E$ \\
    \hline
    2D FS + 2D phonon ($z_c=2$) & $T^2\ln \frac{T_\text{FL}}{T}$ & $T^{\frac{2}{z_p-1}-1}$ & $T^{\frac{2}{z_p}}$ & $T$ & $T$ \\
    \hline
    2D FS + 3D phonon ($z_c=3$) & $T^2\ln \frac{T_\text{FL}}{T}$ & $T^{\frac{2}{z_p-1}}$ & $T^{\frac{3+z_p}{2z_p}}$ & $T$ & $T$ \\
    \hline
    3D FS + 3D phonon ($z_c=3$) & $T^2$ & $T^{\frac{4}{z_p-1}-1}$ & $T^{\frac{3}{z_p}}$ & $T$ & $T$ \\
    \hline
  \end{tabular}
  \caption{Leading temperature dependence of the electron-phonon scattering rate $\Gamma_\text{ep}$ in the crossover regimes introduced in Fig.~\ref{fig:selfenergy1}. For a given scenario, regimes (A,B,C,D) apply when $1\le z_p\le z_c$, whereas for $z_p>z_c$ the regimes (B,C,D) collapse into the single broadened equipartition-like regime (E). Here $z_c=d$ is set by the spatial dimensionality of the phonon dispersion.}
  \label{tab:selfenergy1}
\end{table}

\subsubsection{Two-dimensional Fermi surface coupled to two-dimensional phonons}

As the first example, let us consider the case of a two-dimensional electronic Fermi surface coupled to a two-dimensional optical phonon. This case is experimentally relevant for two-dimensional materials tuned towards electronic criticality. To be concrete, we consider the phonon dispersion, 
\begin{equation}\label{eq:omega2D2D}
  \omega_\vec{q}^2=\omega_D^2+C^2\vn{q_\text{2D}}^{z_p-1}\,,
\end{equation} where $\vec{q}_\text{2D}$ is the projection of $\vec{q}$ on to the two-dimensional (XY) plane, and there is no $q_z$ dispersion. The Landau-damping parameters of the phonon are computed to be $\gamma=\gamma_\text{2D}=2\calN_\text{2D}\lambda^2$, and $\eta_\vec{q}=v_F\vn{q_\text{2D}}$; see Eq.\ref{eq:Delta_Pi_X}.

Let us introduce the relevant energy scales associated with the system. The first two are clearly the temperature $T$, and the renormalized Debye frequency $\omega_D$, respectively. The third scale is the effective phonon bandwidth defined by 
\begin{equation}\label{eq:omega_Lambda}
  \omega_\Lambda\sim C \Lambda^{\frac{z_p-1}{2}}\,,
\end{equation} where $C$ is the dispersion coefficient introduced in Eq.~\eqref{eq:omega2D2D}, and $\Lambda\sim 2k_F$ is the phonon momentum cutoff determined by the maximum allowed backscattering due to the electronic Fermi surface. The final energy scale $\omega_*$ involves the damping coefficient $\gamma$:
\begin{equation}\label{eq:omega_star}
  \omega_*\sim C^{2/(z_p+1)}(\gamma/v_F)^{(z_p-1)/(z_p+1)}\,.
\end{equation} 
This energy scale follows from dimensional analysis, noting that $\gamma$ has dimensions of $[\text{Energy}]^2$, but appears in the computation in the form of $\gamma/v_F$. We note that $\omega_*$ is a unique energy scale that can appear at low-energy (excluding $\omega_\Lambda$), and organizes the crossover diagram as shown in Fig.~\ref{fig:selfenergy1}. The scale $\omega_*$ reflects the competition between two effects: Phonon dispersion and Landau damping. When the phonon energy is smaller than $\omega_*$, it is overdamped and in the opposite limit it is underdamped. Let us now discuss the characteristic features associated with each of the regimes labeled as A-E in Fig.~\ref{fig:selfenergy1}.

\begin{itemize}
    
 \item[(A)] In the Fermi-liquid regime, the electron-phonon scattering rate, 
 \begin{equation}\label{eq:Gammaep_A_2D}
 \begin{split}
   \Gamma_\text{ep,A}&\sim \frac{\gamma^2}{\omega_D^4}\frac{T^2}{k_F v_F} \ln \frac{T_\text{FL}}{T}\,,
 \end{split} 
 \end{equation} where the temperature scale $T_\text{FL}$ is given by, 
 \begin{equation}\label{eq:TFL}
   T_\text{FL}\sim v_F \left(\frac{\omega_D}{C}\right)^{\frac{2}{z_p-1}} \omega_D^2/\gamma\propto \omega_D^{2z_p/(z_p-1)}\,.
 \end{equation}
 Here, we have kept only the leading $T$-dependent term and dropped unimportant numerical prefactors. In this regime, most phonons are not thermally excited but they serve as virtual gapped quasiparticles that mediate an effective electronic interaction, leading to a FL scaling of $\Gamma_\text{ep}$. The $T^2\ln(1/T)$ behavior is a well-known feature of two-dimensional FL \cite{AVChubukov2003}. In the $(\omega_D,T)$-plane, this regime is defined by the condition 
 \begin{equation}\label{}
   T\ll \min(T_\text{FL},\omega_D)\,.
 \end{equation} Here, the former inequality applies when $\omega_D\ll \omega_*$ (overdamped phonons), and the latter applies when $\omega_D\gg \omega_*$ (underdamped phonons).

\item[(B)] The generalized Bloch-Gr\"uneisen regime is adiabatically connected to the usual Bloch-Gr\"uneisen regime, except that the phonon dispersion we consider here is different. In this regime, the main contribution to the electron-phonon scattering rate is from the thermally excited, underdamped phonons, yielding
 \begin{equation}\label{eq:Gammaep_B1}
   \Gamma^{z_p<2}_\text{ep,B}\sim \frac{\gamma }{k_F}\frac{T^{\frac{3-z_p}{z_p-1}}}{C^{\frac{2}{z_p-1}}}\,.
 \end{equation}  When $z_p=2$, the power-law receives additional logarithmic corrections, 
 \begin{equation}\label{eq:Gammaep_B2}
   \Gamma_\text{ep,B}^{z_p=2}\sim \frac{\gamma T}{k_F C^2} \ln\frac{T}{\omega_D}\,.
 \end{equation}
 
 The crossover boundary of this regime is set by the condition 
 \begin{equation}\label{}
   \max(\omega_*,\omega_D)\ll T \ll \sqrt{\omega_D^2+\omega_\Lambda^2}\,.
 \end{equation} The left inequality applies when some of the phonons are thermally excited but not overdamped, while the right inequality applies when not all phonons in the band are excited.  We also note that when $z_p<2$, $\Gamma_\text{ep}$ is independent of $\omega_D$ at the leading order, and $\omega_D$ is an irrelevant correction in the scaling sense. 
 

 \item[(C)] In the electron-phonon critical regime, the main contribution to electron-phonon scattering arises from the thermally excited overdamped phonons. The resulting electron-phonon system behaves as a critical non-Fermi liquid, where the phonon plays the role of the critical boson with,
 \begin{equation}\label{eq:Gammaep_C1}
    \Gamma_\text{ep}^{z_p<2}\sim \frac{v_F}{k_F} \left(\frac{\gamma T}{v_F C^2}\right)^{\frac{2}{z_p}}\,,
 \end{equation} Here the $T^{2/z_p}$ scaling agrees with the non-Fermi liquid expectation \cite{DFMross2010,ZDShi2022,ZDShi2023a,HGuo2024d}. When $z_p=2$, the power-law again leads to logarithmic corrections: 
 \begin{equation}\label{eq:Gammaep_C2}
   \Gamma_\text{ep,C}^{z_p=2}\sim \frac{\gamma T}{k_F C^2} \ln \frac{C^2 \gamma T}{v_F\omega_D^4}\,.
 \end{equation}

 The crossover boundary of the regime is given by 
 \begin{equation}\label{}
   T_\text{FL} \ll T \ll \omega_*\,.
 \end{equation} Here the above condition means that the temperature should be high enough so that there is a substantial density of thermally excited phonons, but the phonon energy cannot be too high so most of the phonons are still overdamped. 
 
\item[(D)] The equipartition regime: When the temperature $T$ is larger than the phonon bandwidth, the system enters the equipartition regime with linear-in-$T$ resistivity. Here $\Gamma_\text{ep}$ is 
\begin{equation}\label{eq:Gammaep_D1}
   \Gamma_\text{ep,D}^{z_p<2}\sim \frac{\gamma T}{k_F} \min\left(\frac{\Lambda^{2-z_p}}{C^2},\frac{\Lambda}{\omega_D^2}\right)\,,
 \end{equation} where $\Lambda\sim 2k_F$ is the UV momentum cutoff. When $z_p=2$, the result becomes logarithmic in $\Lambda$
 \begin{equation}\label{eq:Gammaep_D2}
   \Gamma_\text{ep,D}^{z_p=2} \sim \frac{\gamma T}{k_F C^2}\ln\frac{\omega_D^2+C^2 \Lambda}{\omega_D^2}\,.
 \end{equation} The temperature range of this regime is set by 
 \begin{equation}\label{}
   T\gg \sqrt{\omega_D^2+\omega_\Lambda^2}\,.
 \end{equation}
 
 \item[(E)] The regimes (B,C,D) discussed above only apply to $z_p\leq z_c=2$. In these three regimes, phonons at higher energy contribute more to the scattering rate. The situation changes when we consider $z_p>z_c=2$, where now the scattering is dominated by the lowest energy phonons. The electron-phonon scattering rate is   

\begin{equation}\label{eq:Gammaep_E}
   \Gamma_{\text{ep,E}}^{z_p>2}\sim \frac{\gamma T}{k_F} \frac{1}{\omega_D^2} \min\left(\left(\frac{\omega_D}{C}\right)^{\frac{2}{z_p-1}},\Lambda\right)\,.
 \end{equation} Here, the main contribution to scattering arises from the phonons at the bottom of the band, and the phase space of these phonons are estimated as $C\vn{q_\text{2D}}^{z_p-1}\leq \omega_D^2$. We also note that the coefficient of the linear-in-$T$ scattering rate diverges when $\omega_D\to 0$. We note that the strict $\omega_D\to0$ limit is  unphysical, because in taking the limit we have ignored the thermal mass of the phonon, which should be self-consistently generated \cite{AJMillis1993,JADamia2020,IEsterlis2021}. Therefore, the result for regime (E) should only apply to values of $\omega_D$ greater than the generated thermal mass. 
 
  This linear-in-$T$ regime (E) now absorbs the previous (B,C,D) regimes, and the temperature of the regime is then 
  \begin{equation}\label{}
    T\gg\max(T_\text{FL},\omega_D)\,,
  \end{equation} as shown in Fig.~\ref{fig:selfenergy1} (b).

 \end{itemize}
 
 Finally, let us comment on the $\omega$-dependence of $\Gamma_\text{ep}$, with more details in Appendix.~\ref{sec:Gamma_omega}. We found that in regimes (A), (B) and (C), the $\omega$-dependence of $\Gamma_\text{ep}$ is parametrically similar to its $T$-dependence, i.e. the scaling can be obtained by replacing $T\to \omega$. However, regime (B) has a different threshold $z_c$, which is given by $z_c^\omega=2d-1$. When $z_p>z_c$, the frequency dependence becomes singular in the IR, which will be cutoff by $\omega_D$. In regime (D), the frequency dependence saturates, while the $T$-dependence can still grow due to the thermally populated phonons, as a result, the frequency-dependence is only a small correction to the linear-in-$T$ piece. Finally, in regime (E), the $T$-dependence and the $\omega$-dependence becomes unrelated: The $T$-dependent piece grows linearly in $T$, and the $\omega$-dependence switches between power-law in $\omega$ or a constant, depending on $z_p\gtrless z_c^\omega$. The same observation applies to the other two scenarios with different dimensionalities.

 \subsubsection{Two-dimensional Fermi surface coupled to three-dimensional phonons}
 
 The second scenario we discuss is when the electrons (and the Fermi surface) are still quasi-2D, but the phonons acquire a  dispersion in the $z$-direction. The renormalized phonon dispersion is 
 \begin{equation}\label{}
   \omega_\vec{q}^2=\omega_D^2+C^2\vn{q_\text{2D}}^{z_p-1}+c_z^2 q_z^2\,.
 \end{equation} Here, the $z$-directional momentum is bounded by $|q_z|\leq \Lambda_z$, where $\Lambda_z=\pi/a_z$ is related to the $z$-directional lattice constant $a_z$. When $c_z \Lambda_z\ll \omega_D$, the $z$-dispersion does not modify $\Gamma_\text{ep}$ significantly from the pure two-dimensional scenario, and we therefore consider the opposite limit $c_z \Lambda_z\gg \omega_D$.
 
 The physics of this scenario is similar to the previous one, and the crossover boundaries have the same parametric dependence, as shown in Fig.~\ref{fig:selfenergy1}. The difference from the first scenario is that the $T$-dependence of $\Gamma_\text{ep}$ is different in regimes (B,C), as the scattering rate in these regimes depend on the phase space volume of the excited phonons, which in turn depends on the dispersion. Another difference is that 3D phonon changes the critical value $z_c$ of the linear-in-$T$ regime (E) to $z_c=3$. We state the results for $\Gamma_\text{ep}$ below, and defer the derivation to Appendix~\ref{sec:funcI}.

 \begin{itemize}
 \item[(A)] In the FL regime, $\Gamma_\text{ep}$ still has the $T^2\ln(T_\text{FL}/T)$ dependence as the Landau damping still arises from the quasi-2D Fermi surface,
 \begin{equation}\label{}
  \Gamma_\text{ep,A} \sim \frac{\gamma^2}{\omega_D^4} \frac{\omega_D}{c_z \Lambda_z} \frac{T^2}{k_F v_F} \ln\frac{T_\text{FL}}{T}\,,
\end{equation}  where the extra dimension of the phonon only modifies the prefactor of $\Gamma_\text{ep}$. 

\item[(B)] In the generalized Bloch-Gr\"uneisen regime, $\Gamma_\text{ep}$ depends on $T$ with a different exponent:
\begin{equation}\label{}
  \Gamma_\text{ep,B}^{z_p<3}\sim \frac{\gamma}{k_F c_z \Lambda_z} \left(\frac{T}{C}\right)^{\frac{2}{z_p-1}}\,. 
\end{equation}  This is one power higher than the counterpart of the first scenario, which can be understood from the fact that the $q_z$-integral multiplies the effective phase space volume by a factor of $T/c_z$. 
 When $z_p=z_c=3$, the result becomes logarithmic, yielding 
\begin{equation}\label{}
  \Gamma_\text{ep,B}^{z_p=3} \sim \frac{\gamma}{k_F c_z \Lambda_z} \frac{T}{C}\ln\frac{T}{\omega_D}\,.
\end{equation}

\item[(C)] In the electron-phonon critical regime, $\Gamma_\text{ep}$ is \
\begin{equation}\label{}
  \Gamma_\text{ep,C}^{z_p<3}\sim \frac{\gamma T}{k_F c_z \Lambda_z} \frac{1}{C}q_1^{\frac{3-z_p}{2}} \sim \frac{\gamma T}{k_F c_z \Lambda_z} \frac{1}{C}\left(\frac{\gamma T}{v_F C^2}\right)^{\frac{3-z_p}{2z_p}}\,.
\end{equation} When $z_p=3$, the integral is logarithmic, yielding 
\begin{equation}\label{}
  \Gamma_\text{ep,C}^{z_p=3}\sim \frac{\gamma T}{k_F c_z \Lambda_z}\frac{1}{C} \ln\left(\frac{\gamma C T}{v_F \omega_D^3}\right)\,.
\end{equation}
 
 \item[(D)] In the equipartition regime, we obtain
 \begin{equation}\label{}
  \Gamma_\text{ep,D}^{z_p<3} \sim \frac{\gamma T}{k_F} \frac{\Lambda^{\frac{3-z_p}{2}}}{C c_z \Lambda_z}\,,
\end{equation} and
\begin{equation}\label{}
  \Gamma_\text{ep,D}^{z_p=3} \sim \frac{\gamma T}{k_F c_z \Lambda_z C} \ln \frac{C \Lambda}{\omega_D}\,.
\end{equation} 

\item[(E)] The results for (B,C,D) above apply when $z_p\leq z_c=3$. When $z_p>z_c=3$, the regions (B,C,D) merge into a single equipartition regime with,
\begin{equation}\label{}
  \Gamma_\text{ep,E}^{z_p>3}\sim \frac{\gamma T}{k_F c_z \Lambda_z} \frac{1}{\omega_D} \left(\frac{\omega_D}{C}\right)^{\frac{2}{z_p-1}}\,.
\end{equation}
\end{itemize}

\subsubsection{Three-dimensional Fermi surface coupled to three-dimensional phonons}

The final scenario we consider here is when both the electrons and the phonons have a three-dimensional dispersion. For simplicity, we assume the Fermi surface is spherical and the phonon dispersion is isotropic,
\begin{equation}\label{}
  \omega_{\vec{q}}^2=\omega_D^2+C^2 \vn{q}^{z_p-1}\,.
\end{equation} The regimes and their crossovers are also described by Fig.~\ref{fig:selfenergy1} (a),(b). 

The electron-phonon scattering rate in the regimes are as follows:
\begin{itemize}
\item[(A)] In the FL regime, $\Gamma_\text{ep}$ is 
\begin{equation}\label{}
  \Gamma_\text{ep,A} \sim \frac{\gamma^2 T^2}{k_F^2 v_F} \begin{cases}
                                                         \frac{1}{\omega_D^4} \left(\frac{\omega_D}{C}\right)^{\frac{2}{z_p-1}}, &  z_p>3/2 \\
                                                         \frac{\Lambda^{3-2z_p}}{C^4}, & 1<z_p<3/2
                                                       \end{cases}
\end{equation} Here, $\Gamma_\text{ep}$ is proportional to $T^2$ as expected from the FL theory, and there is no logarithmic when the fermion dispersion is 3D. The coefficient of the $T^2$ term behaves differently when $z_p\gtrless 3/2$. When $z_p<3/2$, the scattering is dominated by phonons with large momentum, and in the opposite limit the scattering is dominated by phonons with small momentum. 

\item[(B)] In the generalized Bloch-Gr\"uneisen regime, we obtain 
\begin{equation}\label{}
  \Gamma_\text{ep,B}^{z_p<3} \sim \frac{\gamma}{k_F^2} \frac{T^{\frac{4}{z_p-1}-1}}{C^{\frac{4}{z_p-1}}}\,.
\end{equation} When $z_p=3$, the result is logarithmic 
\begin{equation}\label{}
  \Gamma_\text{ep,B}^{z_p=3}\sim \frac{\gamma T}{k_F^2 C^2} \ln\frac{T}{\omega_D}\,. 
\end{equation}

\item[(C)] In the electron-phonon critical regime, we obtain 
\begin{equation}\label{}
  \Gamma_\text{ep,C}^{z_p<3} \sim \frac{v_F}{k_F^2} \left(\frac{\gamma T}{v_F C^2}\right)^{\frac{3}{z_p}}\,,
\end{equation}
\begin{equation}\label{}
  \Gamma_{\text{ep,C}}^{z_p=3} \sim \frac{\gamma T}{k_F^2 C^2} \ln \left(\frac{\gamma C}{v_F \omega_D^2}\frac{T}{\omega_D}\right)\,.
\end{equation} Here, the $T^{3/z_p}$ exponent also agrees with 3D non-Fermi liquid \cite{BLAltshuler1994}.

\item[(D)] In the equipartition regime, we obtain 
\begin{equation}\label{}
  \Gamma_\text{ep,D}^{z_p<3} \sim \frac{\gamma T}{k_F^2 C^2} \Lambda^{3-z_p}\,,
\end{equation}
\begin{equation}\label{}
  \Gamma_\text{ep,D}^{z_p=3} \sim \frac{\gamma T}{k_F^2 C^2} \ln \frac{C \Lambda}{\omega_D}\,.
\end{equation}

The regimes (B,C,D) apply to $z_p\leq z_c=3$, and when $z_p>z_c=3$, the equipartition regime (E) absorbs the regimes (B,C,D), and we obtain 
\begin{equation}\label{}
  \Gamma_\text{ep,E}^{z_p>3}\sim \frac{\gamma T}{k_F^2}\frac{1}{\omega_D^2} \left(\frac{\omega_D}{C}\right)^{\frac{4}{z_p-1}}\,.
\end{equation}
\end{itemize}

\subsection{Transport scattering rate}\label{sec:umklapp}

We now discuss under what conditions the single-particle scattering rate $\Gamma_\text{ep}$ obtained in Sec.~2.2 can be used as a proxy for the transport scattering rate $\Gamma_\text{tr}$. Two additional ingredients are essential for this step: momentum relaxation, so that the phonons can act as an effective momentum sink, and large-angle scattering, so that electron-phonon collisions efficiently relax the electrical current. In the present setting, these requirements are provided by a combination of umklapp scattering and weak disorder.

First, some mechanism must break exact momentum conservation. If the combined electron-phonon system conserved momentum exactly, phonon drag would lead to an infinite dc conductivity \cite{YGGurevich1989}. Throughout this section, we therefore assume that the phonon system can exchange momentum efficiently with other degrees of freedom---for example through disorder or through coupling to other phonon modes---so that phonon drag does not dominate the transport response.

Second, one must overcome the usual small-angle suppression of transport scattering. If only normal electron-phonon processes are retained, then the transport scattering rate is reduced relative to the single-particle rate by the familiar angular factor $\delta\theta^2$ \cite{GDMahan2000}, where $\delta\theta\sim |\vec q|/k_F$ is the typical change in the electron momentum direction induced by a phonon of momentum $\vec q$. As a result, $\Gamma_\text{tr}$ is parametrically smaller than $\Gamma_\text{ep}$. By contrast, once phonon-mediated umklapp processes are available, the scattering can involve large momentum transfer and the transport rate becomes parametrically comparable to the single-particle rate \cite{PALee2021}.

The relevant momentum configuration is shown in Fig.~\ref{fig:umklapp_config}. An electron with momentum $\vec k$ on the FS can equivalently be viewed as a state with momentum $\tilde{\vec k}=\vec k+\vec b$ on the FS repeated in a neighboring Brillouin zone, where $\vec b$ is a reciprocal lattice vector. A phonon with momentum $\vec q$ can then scatter this state to another point $\vec k'=\tilde{\vec k}+\vec q$ on the neighboring FS. The minimal phonon momentum required to connect the two FSs is denoted by $\Delta_q$. Phonons with $|\vec q|<\Delta_q$ do not participate in umklapp scattering, while those with $|\vec q|\gtrsim \Delta_q$ can efficiently relax the current.

Weak disorder also helps mitigate the familiar hot-spot bottleneck \cite{ARosch1999,PALee2021}. In a clean system, the dominant umklapp scattering occurs only on portions of the FS near the Brillouin-zone boundary, leaving other portions comparatively ``cold.'' Transport can then be short-circuited by these cold regions. Weak impurity scattering alleviates this problem by mixing momenta around the FS and transferring carriers from the cold regions into the hot regions where umklapp scattering is active.

The momentum threshold $\Delta_q$ introduces an additional infrared cutoff into the transport problem. The softest phonons near $\vec q=0$, which dominate the single-particle scattering rate discussed in Sec.~2.2, are no longer available for current relaxation unless their momentum exceeds $\Delta_q$. The appropriate low-energy scale entering the umklapp transport rate is therefore
\begin{equation}
    \omega_\text{eff}=\max\!\left(\omega_D,\; C\,\Delta_q^{(z_p-1)/2}\right),
\end{equation}
which replaces $\omega_D$ as the effective infrared cutoff in the scaling analysis of phonon-mediated transport. With this substitution, the parametric temperature dependence of the transport scattering rate follows the results of Sec.~2.2. In particular, once umklapp is operative and weak disorder suppresses phonon drag and hot-spot short-circuiting, $\Gamma_\text{tr}$ inherits the same parametric temperature dependence as $\Gamma_\text{ep}$, with the low-energy crossover scales controlled by $\omega_\text{eff}$ rather than $\omega_D$.

\usetikzlibrary{decorations.pathreplacing}
\begin{figure}
  \centering
  \begin{tikzpicture}
    \draw[dotted] (0,0) rectangle (100pt,100pt);
    \draw[dotted] (100pt,0) rectangle (200pt, 100pt);
    \draw[thick] (50pt,50pt) circle (40pt);
    \draw[thick] (150pt,50pt) circle (40pt);
    \filldraw[fill=black] (150pt,50pt) circle (2pt);
    \draw[decorate,decoration={brace},thick]   (110pt,48pt)--(90pt,48pt) node[midway,below] {$\Delta_q$}; 
    \draw[thick,->] (150pt,50pt)--(190pt,50pt);
    \node[below] at (180pt,50pt) {$\vec{k}$};
    \draw[thick,->] (150pt,50pt)--(90pt,50pt) node [near start,below=2pt] {$\vec{\tilde{k}}=\vec{k}+\vec{b}$};
    \draw[thick,->] (150pt,50pt)--(112.412pt,63.6808pt) node [near start,above=10pt,xshift=3pt] {$\vec{k'}=\vec{\tilde{k}}+\vec{q}$};
    \draw[thick,->]  (90pt,50pt)--(112.412pt,63.6808pt) node[near end,below] {$\vec{q}$};
    \draw[thick,->] (150pt,90pt)--(50pt,90pt) node[pos=0.4,above=-3pt] {$\vec{b}$};
    \node[below=3pt] at ($(50pt,50pt)+(-40:40pt)$) {FS};
    \node[below=3pt] at ($(150pt,50pt)+(-40:40pt)$) {FS};
  \end{tikzpicture}
  \caption{Momentum configuration for phonon-mediated umklapp scattering. The two circles denote the same FS repeated in adjacent Brillouin zones, separated by a reciprocal lattice vector $\vec b$. A fermion with momentum $\vec k$ on the right FS is equivalently described by $\tilde{\vec k}=\vec k+\vec b$ on the neighboring FS, and can be scattered to $\vec k'=\tilde{\vec k}+\vec q$ by exchanging a phonon of momentum $\vec q$. The minimal phonon momentum needed to connect the two FSs is $\Delta_q$. The FS is drawn to be circular for simplicity.}
  \label{fig:umklapp_config}
\end{figure}
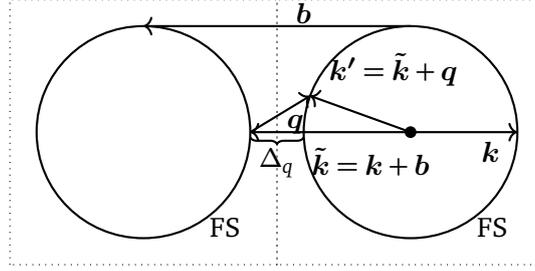

\section{Phonon softening from electronic criticality}\label{sec:softening}

In Section~\ref{sec:electronphonon} we established a general criterion for softened optical phonons to generate asymptotic low-temperature $T$-linear scattering: after Landau damping is included, the phonon must acquire a dynamical exponent $z_p>d$. The purpose of the present section is to test that criterion in a concrete microscopic setting. We study a symmetry-allowed nonlinear coupling between an optical phonon and a $\vec Q=0$ electronic collective mode, which is motivated by THz pump-probe experiments that studied energy relaxation in the cuprates \cite{DChaudhuri2025}, and first proposed theoretically in Ref.~\cite{Guo2025}, and ask how strongly the phonon is softened as the system approaches the electronic quantum critical point.

Our strategy is to proceed in two steps. We first compute the phonon self-energy generated by the critical electronic mode and extract the resulting renormalized phonon dispersion, temporarily neglecting the feedback of the softened phonon on the electronic critical sector. We then compare the resulting $z_p$ with the criterion derived in Sec.~\ref{sec:electronphonon}, and later return to the effects of feedback. In the clean theory, scenarios (i) and (iii) (where the electronic sector and the phonon sector share the same spatial dimension) place the system at the marginal boundary $z_p=d$, so the mechanism is at best borderline for asymptotic $T$-linear transport. By contrast, scenario (ii) is less favorable: the electronic sector still yields $z_p=2$, but the relevant threshold from Sec.~\ref{sec:electronphonon} is $z_c=3$, so the system remains on the multi-crossover side rather than entering the broadened equipartition regime. Including feedback generally tends to reduce the effective $z_p$ in (i) and (iii) further, moving the system away from low-temperature $T$-linear scattering.

\subsection{The Model} \label{sec:model_softening}

We consider an imaginary-time Lagrangian for a metallic Fermi liquid coupled both to a quantum critical collective mode and to optical phonons,
\[
\calL=\calL_e+\calL_p+\calL_{ep}+\delta\calL_e+\delta\calL_{ep},
\]
where $\calL_e,\calL_p,\calL_{ep}$ were introduced in Eq.~\eqref{eq:action1}, and we now make the corresponding interaction terms explicit by specifying the electron-electron interaction and the additional phonon-critical-mode coupling:
\begin{subequations}\label{eq:action}
\begin{eqnarray}\label{eq:Le}
    \delta\calL_e & =& \int \rd\tau \Big[\sum_{\vec{q},\alpha} \varphi_{\alpha,-\vec{q}}\left(-\partial_\tau^2+v_{\vp}^2\vec{q}^2+ r\right)\varphi_{\vec{q},\alpha} \nn \\
  &&  + g \sum_{\vec{k},\vec{q},\sigma,\sigma',\alpha} f_{\vec{k},\vec{{q}},\alpha}^{\sigma\sigma'} c_{\vec{k+q/2},\sigma}^\dagger c^{\phantom\dagger}_{\vec{k-q/2},\sigma'}\varphi_{\alpha,\vec{q}}\Big], \,\\
\label{eq:Lep}\delta \calL_{ep} &=&  \frac{u}{2}\int \rd \tau \sum_{\vec{k},\vec{{q}},a,\alpha,\beta} L^{a}_{\vec{k},\vec{q},\alpha\beta} \varphi_{\vec{-k+q/2},\beta} \varphi_{\vec{k+q/2},\alpha} X_{a,-\vec{q}}.\,
\end{eqnarray}
\end{subequations}
Beyond the Lagrangian introduced in Eq.~\eqref{eq:action1}, Eq.~\eqref{eq:action} adds a Yukawa coupling between the electrons and the long-wavelength fluctuations of the bosonic collective mode $\varphi$. The field $\varphi$ is precisely the order parameter associated with the $\vec Q=0$ quantum phase transition under consideration, and criticality is reached by tuning $r$ so that the renormalized mass $m_\varphi^2=r-r_c$ vanishes. Depending on the scenario of interest, $\varphi$ disperses only in 2D [scenarios (i) and (ii), where $\vec q^2\to \vec q_\text{2D}^{\,2}$] or in 3D [scenario (iii)].

It is useful to distinguish the roles of the three couplings $g$, $\lambda$, and $u$. The Yukawa coupling $g$ controls the critical electronic dynamics associated with $\varphi$. The conventional deformation-potential coupling $\lambda$ is the coupling that enters the transport analysis of Sec.~\ref{sec:electronphonon} and generates the standard Landau damping of the phonon. By itself, however, $\lambda$ does not make the static phonon renormalization parametrically sharper as the electronic critical point is approached; to leading order within Eliashberg theory, the corresponding static phonon self-energy is the familiar $m_\varphi$-independent shift $\sim \calN \lambda^2$ \cite{AVChubukov2020}. By contrast, the nonlinear coupling $u$ directly couples the phonon to the critical boson $\varphi$, so the resulting phonon self-energy becomes explicitly sensitive to the distance $m_\varphi$ from the electronic quantum critical point and is correspondingly enhanced as criticality is approached. This is the coupling that can generate the additional low-energy phonon softening studied in the present section. \change{It is important here that $X$ denotes an optical phonon coupled to the collective mode $\varphi$, so it is not constrained by Goldstone's theorem or by an acoustic shift symmetry.} Microscopically, the coupling can be generated by fermion mediation. A simple computation in Fermi gas shows that this coupling is proportional to the derivative of the density of states at the Fermi level, as well as the Yukawa coupling $g$ and the conventional electron-phonon coupling $\lambda$, $u\propto\calN'(0)g^2\lambda$ \cite{JAHertz1974}. However, in the present discussion we will treat the coupling $u$ as a free parameter. For the dynamical part of the phonon self-energy, we will assume that the Landau damping induced by $\lambda$ provides the dominant contribution, which is already discussed in Sec.~\ref{sec:electronphonon}. The rationale is that this damping channel couples directly to the electronic particle-hole continuum and is therefore weighted by the full electronic density of states at the Fermi surface.

The nature of the critical point is encoded in the form factor $f_{\vec{k},\vec{q},\alpha}^{\sigma\sigma'}$. Examples include: (a) Ising-nematic criticality in the charge channel, where $\varphi_\alpha=\varphi$ is a scalar and $f_{\vec{k},\vec{q},\alpha}^{\sigma\sigma'}=\delta^{\sigma\sigma'}\cos 2\theta_{\vec{k}}$, with $\theta_{\vec{k}}$ the direction of $\vec{k}$ in 2D; and (b) ferromagnetic criticality in the spin channel, where $\varphi_\alpha=(\varphi_x,\varphi_y,\varphi_z)$ and $f_{\vec{k},\vec{q},\alpha}^{\sigma\sigma'}=\frac{1}{\sqrt{2}}(\sigma^x,\sigma^y,\sigma^z)$, with $\sigma^{x,y,z}$ the Pauli matrices acting on the spin indices.

In Sec.~\ref{sec:softening1}, we analyze the phonon softening under the following simplifying assumptions:
\begin{enumerate}
    \item For explicit estimates, we take the form factor to be $f_{\vec{k},\vec{q},\alpha}^{\sigma\sigma'}=\delta^{\sigma\sigma'}$. This is exact for the ferromagnetic quantum critical point in the symmetric phase, and for the Ising-nematic case it amounts to neglecting the anisotropy of the form factor along the Fermi surface.
    \item Disorder effects are assumed to be weak, so that the impurity scattering rate $\Gamma$ remains small compared with the other scales entering the Eliashberg analysis.
    \item In the first pass through the problem, we neglect the feedback of the softened phonon on the electronic critical sector, so that the critical dynamics of $\varphi$ are controlled by the electron-only theory. We return to this feedback later in the section and discuss how it modifies the conclusion.
\end{enumerate}
We will comment at various points on the consequences of relaxing these assumptions.

\change{Before proceeding, let us discuss other possible couplings between the phonon $X$ and the critical mode $\varphi$. The first is a possible linear coupling between $\varphi$ and $X$. Such a coupling is not universal; for example, it is forbidden by time-reversal symmetry when $\varphi$ is a ferromagnetic order parameter, while it is allowed for Ising-nematic order. We do not include it in the main analysis because the mechanism of interest here is the nonlinear coupling in Eq.~\eqref{eq:Lep}, which makes the phonon self-energy explicitly sensitive to proximity to the electronic critical point. A potential concern is that a linear coupling to the lattice may qualitatively affect the low-energy critical fluctuations, as in the case of acoustic phonons \cite{IPaul2017}. For acoustic phonons, Goldstone's theorem requires the displacement field to couple derivatively. If a linear coupling between $\varphi$ and $\partial X$ exists, then integrating out the acoustic phonon generates a $\varphi$ self-energy that starts at order $\vec q^{\,2}$, with an explicit dependence on the direction of $\vec q$. This can lead to direction-selective criticality, where $\varphi_{\vec q}$ becomes gapless only along specific directions. The optical case considered here is different because the optical phonon is not a Goldstone mode. If a linear $\varphi$-$X$ coupling is present, its main effect is to hybridize two non-Goldstone modes and shift the resulting normal-mode dispersions, rather than to introduce the specific criticality-enhanced phonon softening analyzed below. Equivalently, integrating out a noncritical optical phonon produces a leading momentum-independent contribution to the $\varphi$ self-energy. The directional dependence appears only in subleading analytic terms, whose scale of variation is expected to be set by microscopic lattice physics, such as the Brillouin-zone size, so it's effect should be weak near the $\Gamma$ point.}

\change{Regarding couplings with higher nonlinearities in $X$ and $\varphi$, the usual infrared power counting makes them subleading to the lowest nonlinear term $X\varphi^2$ considered here. On the one hand, $\varphi$ has a positive engineering dimension, so couplings with higher powers of $\varphi$ generate renormalized phonon dispersions that are no more relevant than the one induced by $X\varphi^2$. On the other hand, since $X$ is gapped, higher powers of $X$ produce only analytic corrections, which are also less relevant than the $X\varphi^2$ contribution considered below.
}

\subsection{Eliashberg analysis}\label{sec:softening1}

We now compute the phonon self-energy generated by the critical electronic mode $\varphi$. Our treatment is based on the Migdal-Eliashberg equations reviewed in Appendix~\ref{sec:YSYK}, which can be made formally controlled in the Yukawa-SYK extension of the model. In this subsection, we keep only the leading contribution in the nonlinear coupling $u$, namely the term of order $u^2$ in the Luttinger-Ward functional, and we  neglect the feedback of the softened phonon on the electronic critical sector. Under these assumptions, the dynamics of $\varphi$ are those of the electron-only Eliashberg theory.

Combining these with the assumptions introduced in Sec.~\ref{sec:model_softening}, the $\varphi$ propagator is dominated by Landau damping from the electronic continuum and takes the form
\begin{equation}\label{eq:Dphi_clean}
  D_{\varphi}^{-1}(i\Omega,\vec{q})=v_\varphi^2\vn{q}^{z_\varphi-1}+m_\varphi^2+\frac{\gamma_g |\Omega|}{v_F\vn{q}}\,.
\end{equation}
Here $\gamma_g=2g^2\calN_\text{2D}$ in 2D and $\gamma_g=\pi g^2\calN_\text{3D}$ in 3D parametrizes the strength of Landau damping. In scenarios (i) and (ii), $\vec q$ should be understood as the in-plane momentum $\vec q_\text{2D}$, whereas in scenario (iii) it denotes the full 3D momentum. Within the electron-only Eliashberg theory one has $z_\varphi=3$, but we write the propagator in the more general form above because later we will discuss effects that can renormalize $z_\varphi$ \cite{CNayak1994,DFMross2010}.

The renormalized phonon dispersion is determined by the static phonon self-energy through
\begin{equation}\label{}
  \omega_\vec{q}^2= \left(\omega_D^0\right)^2+c^2\vn{q}^2-\Pi_X(\Omega=0,\vec{q})\,.
\end{equation}
It is useful to decompose
\[
\Pi_X=\Pi_X^{(\lambda)}+\Pi_X^{(u)},
\]
where the first term comes from the conventional deformation-potential coupling $\lambda$, and the second from the nonlinear coupling $u$. The role of $\Pi_X^{(\lambda)}$ is familiar:
\begin{equation}\label{eq:PiXlambda}
  \Pi_X^{(\lambda)}(i\Omega,\vec{q})=-\lambda^2\int\frac{\rd \omega}{2\pi} \sum_{\vec{k}} G(i\omega+i\Omega,\vec{k}+\vec{q}) G(i\omega,\vec{k})\,.
\end{equation}
At zero external frequency and to leading order within Eliashberg theory, this gives the standard static shift
\[
\Pi_X^{(\lambda)}(\Omega=0,\vec{q})=\calN \lambda^2,
\]
which is insensitive to the distance $m_\varphi$ from the electronic critical point. We therefore absorb this contribution into the bare optical gap $\omega_D^0$. The criticality-sensitive softening comes instead from $\Pi_X^{(u)}$, while the dynamical damping relevant for Sec.~\ref{sec:electronphonon} is still assumed to be dominated by the conventional Landau damping generated by the electron continuum, which is encoded in $\Pi_X^{(\lambda)}(i\Omega,\vec{q})$.

The leading contribution from the nonlinear coupling is
\begin{equation}\label{eq:PiX_main}
\begin{split}
  \Pi_{X}^{(u)}(i\Omega,\vec{q})=\frac{u^2}{2} \int\frac{\rd \nu}{2\pi}\sum_{\vec{p}}  D_\varphi(i\nu,\vec{p})D_{\varphi}(i\nu+i\Omega,\vec{p}+\vec{q}).
\end{split}
\end{equation}
Unlike $\Pi_X^{(\lambda)}$, this term depends explicitly on $m_\varphi$. Away from criticality, $m_\varphi\to\infty$, it vanishes as $1/m_\varphi^4$. At the quantum critical point, $m_\varphi=0$, we obtain
\begin{equation}\label{}
  \Pi_X^{(u)}(i\Omega=0,\vec{q})\sim \frac{u^2}{a_z^{3-d_\text{el}}} \int \rd \nu \rd^{d_\text{el}}\vec{p} \frac{1}{v_\varphi^2\vn{p}^{z_\varphi-1}+\frac{\gamma_g}{v_F}\frac{|\nu|}{\vn{p}}} \frac{1}{v_\varphi^2\vn{p+q}^{z_\varphi-1}+\frac{\gamma_g}{v_F}\frac{|\nu|}{\vn{p+q}}}\,.
\end{equation}
Here $d_\text{el}=2$ or $3$ is the dimension of the electronic sector. Since we normalize couplings relative to the 3D case, the quasi-2D cases carry an additional factor of $1/a_z$, with $a_z$ the interlayer lattice spacing.

The low-energy momentum dependence of this integral can be extracted by power counting. The relevant loop frequencies are set by balancing the two terms in the $\varphi$ propagator,
\[
v_\varphi^2 p^{z_\varphi-1}\sim \frac{\gamma_g}{v_F}\frac{|\nu|}{p},
\]
which gives
\[
|\nu|\sim \frac{v_\varphi^2 v_F}{\gamma_g}\, p^{z_\varphi}.
\]
For the momentum-dependent part of the self-energy we then take $p\sim q$, so the loop measure scales as $\rd \nu\, \rd^{d_\text{el}}p\sim q^{z_\varphi+d_\text{el}}$, while the two propagators together contribute a factor $q^{-2(z_\varphi-1)}$. The resulting net momentum power is therefore
\begin{equation}
\delta=z_\varphi+d_\text{el}-2(z_\varphi-1)=d_\text{el}+2-z_\varphi\,.
\end{equation}
In the cases of interest, $\delta>0$, so the static integral is UV dominated: the leading term is an analytic cutoff-dependent constant, while the nontrivial infrared information is carried by the first $q$-dependent correction. Accordingly,
\begin{equation}\label{}
  \Pi_X^{(u)}(i\Omega=0,\vec{q})\sim \frac{u^2 v_F}{a_z^{3-d_\text{el}} v_\varphi^2 \gamma_g}\left[\alpha_0 \Lambda^{\delta}-\alpha_1 \vn{q}^{\delta}\right]\,,
\end{equation}
where $\Lambda$ is a UV momentum cutoff and $\alpha_0,\alpha_1$ are positive numerical constants. The momentum-dependent part of this expression determines the effective phonon dynamical exponent,
\begin{equation}\label{eq:zp_res}
  z_p=\delta+1=3+d_\text{el}-z_\varphi\,.
\end{equation}

This result has an important consequence: even exactly at the electronic quantum critical point, the enhancement of the phonon self-energy remains finite. Complete softening of the optical phonon is therefore not automatic. Instead, it requires tuning the nonlinear coupling to a threshold value $u=u_\text{max}$ such that
\[
\Pi_X^{(u=u_\text{max})}(\Omega=0,\vec{q}\to 0)=(\omega_D^0)^2.
\]
The same threshold also marks the stability bound for the theory. If $u>u_\text{max}$, the renormalized phonon frequency vanishes already at a finite distance from the electronic critical point, implying that a structural transition preempts the putative electronic quantum critical point.

The physically interesting regime is therefore $u\lesssim u_\text{max}$, where the optical phonon is strongly softened but the system remains stable. For the saddle-point value $z_\varphi=3$, Eq.~\eqref{eq:zp_res} gives $z_p=d_\text{el}$, so the clean theory lands at or near the marginal cases identified in Sec.~\ref{sec:electronphonon}. We now make this more explicit in 2D and 3D.

For $d_\text{el}=2$ [relevant to scenarios (i) and (ii)], the tuned critical limit $m_\varphi=0$ and $u=u_\text{max}$ gives
\[
\omega_\vec{q}^2 \sim C^2 \vn{q_\text{2D}},
\]
in agreement with $z_p=2$. Moving away from this limit can happen in two distinct ways. Reducing $u$ below $u_\text{max}$ restores a finite optical gap and gives
\[
\omega_\vec{q}^2=\omega_D^2+C^2\vn{q_\text{2D}}.
\]
By contrast, moving away from the electronic critical point at fixed $u$ introduces a crossover controlled by $\vn{q_\text{2D}}/m_\varphi$: for $\vn{q_\text{2D}}\ll m_\varphi$ the dispersion is approximately quadratic,
\[
\omega_\vec{q}^2=\omega_D^2+C_1^2\vn{q_\text{2D}}^2,
\]
whereas for $\vn{q_\text{2D}}\gg m_\varphi$ it crosses over to the critical form
\[
\omega_\vec{q}^2=\omega_D^2+C^2\vn{q_\text{2D}}.
\]

For $d_\text{el}=3$ [scenario (iii)], Eq.~\eqref{eq:zp_res} instead gives $z_p=3$, which means that the leading momentum dependence of $\omega_\vec q^2$ remains quadratic. In other words, the critical mode primarily renormalizes the coefficient of the $q^2$ term rather than producing a softer nonanalytic power. Locally, the dispersion therefore retains the form
\[
\omega_\vec{q}^2=\omega_D^2+C^2\vn{q}^2,
\]
but both $\omega_D$ and the renormalized stiffness $C^2$ depend on $u$ and on the ratio $\vn{q}/m_\varphi$. When $\vn{q}\ll m_\varphi$, the renormalization from $\varphi$ is cut off by its finite mass and $C^2$ is close to its bare value; when $\vn{q}\gg m_\varphi$, the critical renormalization is active and $C^2$ is correspondingly enhanced.

\begin{figure}
    \centering
    \includegraphics[width=0.8\linewidth]{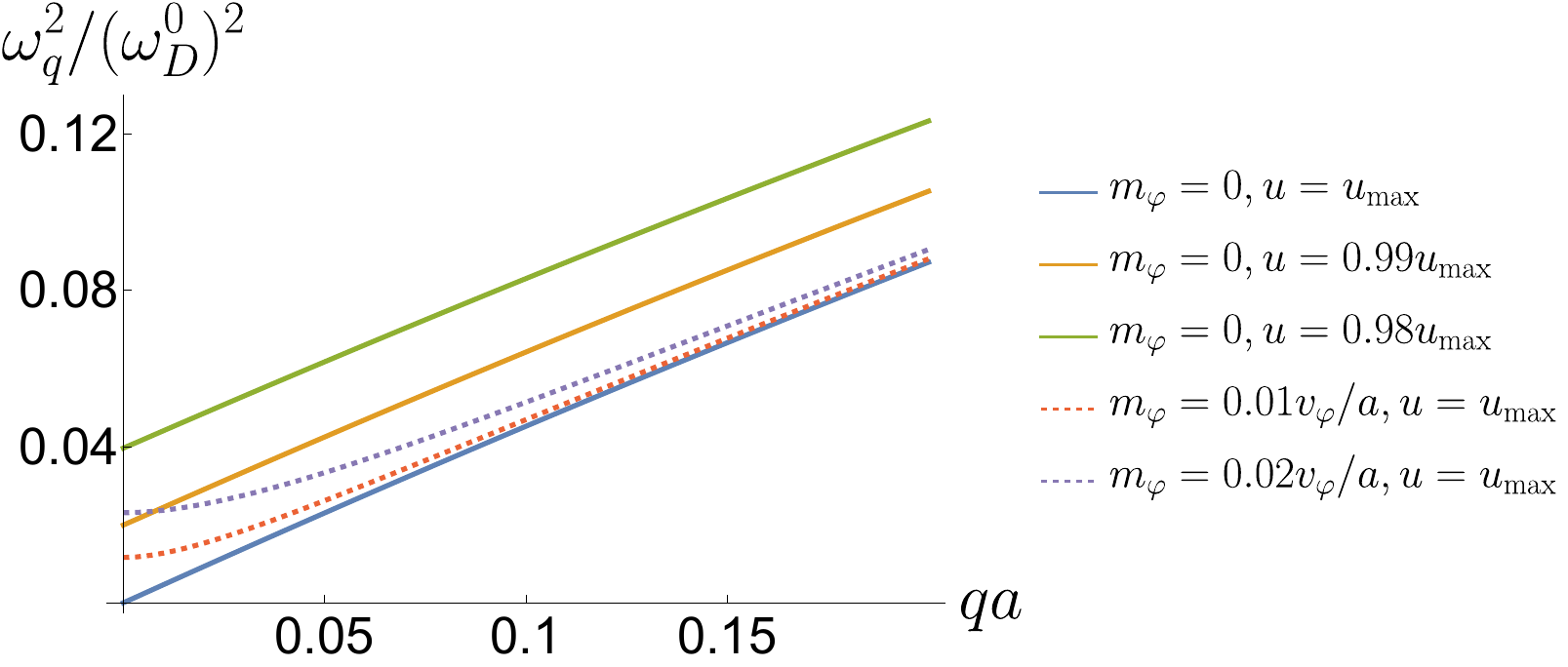}
    \caption{Renormalized optical-phonon dispersion near the $\Gamma$ point for $d_\text{el}=2$. The blue solid curve shows the tuned critical case $m_\varphi=0$ and $u=u_\text{max}$, where the phonon is fully softened and $\omega_{\vec q}^2\propto \vn{q_\text{2D}}$, consistent with the power counting. The other solid curves illustrate the effect of reducing $u$ below $u_\text{max}$ at criticality, which restores a finite optical gap. The dashed curves show the effect of moving away from the electronic quantum critical point at fixed coupling, producing a crossover from quadratic dispersion at the smallest momenta to the critical form at larger $\vn{q_\text{2D}}$.}
    \label{fig:dispersion_Gamma}
\end{figure}

These trends are illustrated in Fig.~\ref{fig:dispersion_Gamma} for the 2D case. The  curve at $m_\varphi=0$ and $u=u_\text{max}$ exhibits the critical form $\omega_{\vec q}^2\propto \vn{q_\text{2D}}$, while moving away from criticality or reducing $u$ restores a finite gap and eventually the conventional quadratic behavior at the smallest momenta.

\subsection{Is there low-temperature linear-in-$T$ resistivity?}

We can now compare the microscopic softening mechanism of Sec.~\ref{sec:softening1} with the general transport criterion derived in Sec.~\ref{sec:electronphonon}. The key input is the effective phonon dynamical exponent $z_p$ generated by the coupling to the critical electronic mode. The overall conclusion is that the clean saddle-point theory is at best marginal for producing asymptotic low-temperature linear-in-$T$ resistivity: scenario (i) lies on the marginal boundary, scenario (ii) is less favorable, and scenario (iii) is again marginal in the leading approximation.

\paragraph{Scenario (i): quasi-2D electrons and 2D phonons.}
In this case we found $z_p=2$. According to Sec.~\ref{sec:electronphonon}, this places the system at the boundary between the broadened equipartition regime of Fig.~\ref{fig:selfenergy1}(b) and the multi-crossover structure of Fig.~\ref{fig:selfenergy1}(a). Correspondingly, $\Gamma_\text{ep}$ crosses over from $T^2\ln(1/T)$ for $T<T_\text{FL}$ to $T\ln(1/T)$ for $T>T_\text{FL}$. Thus the scattering rate is not strictly linear in $T$, but acquires a logarithmic correction. The same conclusion applies to the resistivity once the transport conditions discussed in Sec.~\ref{sec:umklapp} are imposed.

\paragraph{Scenario (ii): quasi-2D electrons and 3D phonons.}
Here the quasi-2D electronic sector still yields $z_p=2$, but the relevant critical value in Sec.~\ref{sec:electronphonon} is now $z_c=3$ because the phonon disperses in 3D. The system therefore remains on the $z_p<z_c$ side, where the scattering rate exhibits the four crossover regimes summarized in Fig.~\ref{fig:selfenergy1}(a). In this case the temperature exponents governing $\Gamma_\text{ep}$ are not close to $1$, so this geometry is less favorable for low-temperature linear-in-$T$ transport.

\paragraph{Scenario (iii): 3D electrons and 3D phonons.}
At the saddle-point value $z_\varphi=3$, Eq.~\eqref{eq:zp_res} gives $z_p=3$. This again places the system on the marginal boundary, now with $z_p=z_c=3$. As in scenario (i), one therefore expects a sequence of crossover regimes in $\Gamma_\text{ep}(T)$ rather than a parametrically broad regime of strict linear-in-$T$ behavior. Depending on the regime, the temperature dependence can be close to linear, but the clean leading-order theory does not guarantee an asymptotically exact $T$-linear resistivity.

The conclusion above is based on the saddle-point result $z_\varphi=3$. A natural question is whether additional effects renormalize $z_\varphi$ and thereby move the system away from marginality. Our expectation is that such corrections tend to increase the effective dynamical exponent of the critical electronic mode, $z_\varphi>3$, which by Eq.~\eqref{eq:zp_res} would reduce $z_p$. If so, the system is pushed further into the regime of Fig.~\ref{fig:selfenergy1}(a), where strict linear-in-$T$ scattering survives only in the conventional high-temperature equipartition regime, while the lower-temperature regimes exhibit exponents larger than $1$. Since the transport scattering rate is bounded from above by the single-particle rate, this would preclude strictly asymptotic linear-in-$T$ resistivity, although the effective exponent could still remain numerically close to $1$ if the deviation of $z_p$ from its marginal value is small.

Two observations support this expectation. First, the leading phonon-feedback calculation in Appendix~\ref{sec:feedback} finds that the feedback is marginal in scenarios (i) and (iii), but irrelevant in scenario (ii), and in the marginal cases it is consistent with an upward renormalization of the effective $z_\varphi$. Second, even within the purely electronic sector, the literature suggests that the saddle-point value $z_\varphi=3$ need not remain exact beyond Eliashberg-type approximations. In Ref.~\cite{MAMetlitski2010}, one finds $z_\varphi=3$ up to three-loop order, where the nonzero diagrams are essentially those already captured by Eliashberg theory. By contrast, Refs.~\cite{THolder2015a,THolder2015} argue that higher-loop fluctuations not included in the Eliashberg equations drive the same qualitative trend toward $z_\varphi>3$. Since those works concern a purely electronic theory, whereas the present problem also involves coupling to the phonon sector, the precise relation between these two effects remains to be understood. Taken together, these considerations suggest that the  leading-order results above should be viewed as an optimistic baseline: even in that limit the system is only marginal for asymptotic low-temperature linear-in-$T$ transport, and additional corrections may push it away from that behavior.

\subsection{Role of disorder and interaction form-factor}

Finally, we revisit the remaining two simplifying assumptions introduced in Sec.~\ref{sec:model_softening}, namely the neglect of weak disorder and the approximation of a constant form factor. Relaxing these assumptions modifies the low-energy phonon softening in different ways. Weak disorder changes the infrared dynamics of the critical boson and can enhance the tendency toward phonon softening near the $\Gamma$ point, while a nontrivial form factor introduces anisotropy and can generate cold regions on the Fermi surface. The purpose of this section is to assess whether either effect qualitatively alters the conclusions of the clean analysis. Unless otherwise stated, we take the clean boson propagator to be given by Eq.~\eqref{eq:Dphi_clean} with $z_\varphi=3$ as the reference point.

\subsubsection{Disorder}\label{sec:disorder}

We introduce weak disorder through an elastic fermion scattering rate $\Gamma$. Disorder affects the problem in two distinct ways. First, it modifies the infrared dynamics of the critical boson $\varphi$, and hence the softening of the optical phonon induced by the nonlinear coupling $u$. Second, it changes the low-temperature electron-phonon scattering rate by cutting off the clean scaling regime at sufficiently small momenta and frequencies.

We begin with the effect on phonon softening. In the disordered regime $v_F\vn{q}\ll \Gamma$, impurity scattering cuts off the ballistic Landau-damping form of the $\varphi$ propagator \cite{HGuo2022a}, which now becomes
\begin{equation}\label{eq:Dphi_disorder}
D_\varphi^{-1}(i\Omega,\vec{q})=v_\varphi^2\vn{q}^{z_\varphi'}+m_\varphi^2+\frac{\gamma_g|\Omega|}{\Gamma}\,.
\end{equation}
In this kinematic regime, the effective dynamical exponent of $\varphi$ is reduced to $z_\varphi'=z_\varphi-1$. The propagator crosses back to the clean form in Eq.~\eqref{eq:Dphi_clean} once $v_F\vn{q}\gg \Gamma$.

Repeating the computation of Eq.~\eqref{eq:PiX_main} with the disorder-modified propagator in Eq.~\eqref{eq:Dphi_disorder}, we find that the momentum power counting becomes
\begin{equation}
\delta=d_\text{el}-z_\varphi'\,.
\end{equation}
Accordingly, the static phonon self-energy acquires the form
\begin{equation}\label{eq:PiXuGamma}
\Pi_X^{(u,\Gamma)}(\Omega=0,\vn{q})\sim \frac{u^2}{v_\varphi^2 a_z^{3-d_\text{el}}}\frac{\Gamma}{\gamma_g}\left[\alpha_0'\left(\frac{\Gamma}{v_\varphi}\right)^\delta-\alpha_1'\vn{q}^\delta\right]\,,
\end{equation}
where $\alpha_0'$ and $\alpha_1'$ are positive numerical coefficients. Combining this with the disorder-modified phonon propagator,
\begin{equation}\label{eq:DX_disorder}
D_X^{-1}(i\Omega,\vec{q})=\omega_\vec{q}^2-\Pi_X(i\Omega,\vec{q})+\frac{\gamma|\Omega|}{\Gamma}\,,
\end{equation}
we see that the effective dynamical exponent of the phonon becomes
\begin{equation}\label{eq:zpp}
z_p'=\delta=d_\text{el}-z_\varphi'\,.
\end{equation}
Thus weak disorder generally makes the low-energy phonon softer than in the clean theory.

This tendency is strongest in 2D. For a local critical boson, $z_\varphi'=2$, the exponent $\delta$ vanishes when $d_\text{el}=2$, and the phonon self-energy acquires a logarithmic enhancement:
\begin{equation}\label{eq:PiXu_disorder}
\Pi_X^{(u)}(\Omega=0,\vn{q})\sim \frac{u^2}{a_z v_\varphi^2}\frac{\Gamma}{\gamma_g} \ln\frac{\Gamma}{\max(m_\varphi,v_\varphi\vn{q_\text{2D}})}\,.
\end{equation}
As the electronic critical point is approached, this logarithm enhances the phonon softening and therefore the tendency toward a nearby structural instability. In the weak-disorder regime, however, this enhancement sets in only at an exponentially small scale. Let $\omega_D$ denote the renormalized optical gap in the disorder-free theory. The disorder-induced correction becomes comparable to $\omega_D^2$ only when
\begin{equation}
m_\varphi\sim \Gamma \exp\left(-\frac{\omega_D^2}{u^2} \frac{a_z v_\varphi^2\gamma_g}{\Gamma}\right)\,.
\end{equation}
For small $\Gamma$, this scale is exponentially suppressed, so in practice the logarithmic enhancement may well be preempted by other low-temperature phenomena before it becomes dominant. In 3D, by contrast, the momentum integral is less infrared sensitive, and the corresponding disorder-induced enhancement of phonon softening is weaker.

We now turn to the consequences for electron-phonon scattering. The clean crossover regimes summarized in Fig.~\ref{fig:selfenergy1} remain valid as long as the typical phonon momentum involved in scattering satisfies $v_F q\gg \Gamma$. Using the clean scaling relation $T\sim q^{z_p}$ for the softened phonon, this condition defines a disorder scale
\begin{equation}
T_\Gamma \sim \frac{v_F C^2}{\gamma} \left(\frac{\Gamma}{v_F}\right)^{z_p}\,.
\end{equation}
For $T\gg T_\Gamma$, the results of Sec.~\ref{sec:electronphonon} continue to apply without modification. For $T\ll T_\Gamma$, however, disorder cuts off the clean low-energy regime. The electron-phonon scattering rate is then further suppressed for two reasons: the available phonon phase space is reduced by a factor of order $v_F\vn{q}/\Gamma$, and the renormalized phonon dispersion crosses over from exponent $z_p$ to the softer disorder-controlled exponent $z_p'$. We elaborate on these two effects in Appendix.~\ref{app:disorder}.

For 2D electrons coupled to a 2D phonon, the clean scaling $\Gamma_\text{ep}\sim T^{2/z_p}$ in the low-temperature regime is  replaced by
\[
\Gamma_\text{ep}\sim T^{2/z_p'},\qquad z_p'=z_p-2.
\]
If the system remains stable, so that $z_p'>0$, this produces a less relevant temperature dependence than in the clean case. For 3D electrons coupled to a 3D phonon, the corresponding low-temperature form becomes
\[
\Gamma_\text{ep}\sim T^{3/z_p'},\qquad z_p'=z_p-2.
\]
In either case, weak disorder does not enlarge the regime of asymptotic linear-in-$T$ scattering; rather, it tends to soften the phonon further while simultaneously introducing a low-temperature cutoff to the clean scaling analysis.

\subsubsection{Form factor}\label{sec:formfactor}

We now briefly comment on the effect of restoring a nontrivial form factor in the electronic sector. As a representative example, consider the Ising-nematic critical point, for which
$
f_{\vec{k}}^{\sigma\sigma'}=\delta^{\sigma\sigma'}\cos 2\theta_{\vec{k}}
$
as in Eq.~\eqref{eq:Le}. Unlike the constant-form-factor approximation used in the main text, this coupling is strongly anisotropic around the Fermi surface and vanishes along the cold-spot directions $\theta_{\vec{k}}=\pi/4+n\pi/2$ with $n=0,1,2,3$ \cite{SAHartnoll2014,VOganesyan2001,XWang2019}.

This anisotropy affects the problem at several levels. First, the fermion scattering rate becomes angle dependent and is suppressed near the cold spots. Second, because the Landau damping of the critical boson $\varphi$ is generated by particle-hole excitations weighted by the same form factor, the damping of $\varphi$ is likewise reduced in those directions. As a result, the phonon self-energy generated by the nonlinear coupling $u$ is no longer controlled by a single isotropic scaling form: the softening of the optical phonon becomes angle dependent, and the low-energy dynamics near the cold spots can differ qualitatively from those in the strongly coupled sectors of the Fermi surface.

A complete treatment of this situation would therefore require a patchwise analysis that keeps the angular structure of both the fermion-boson vertex and the induced phonon self-energy. In particular, the phonon softening itself becomes anisotropic, because the Landau damping of $\varphi$ vanishes along the cold-spot directions. The isotropic exponents derived above should therefore be understood only as an effective characterization of the simplified constant-form-factor problem, rather than as a uniform description of the full anisotropic system.

For the transport problem, however, the conclusions of the previous sections are still expected to capture the dominant contribution. The reason is that the resistivity is controlled primarily by phonon-mediated umklapp processes near the Brillouin-zone boundary, as illustrated in Fig.~\ref{fig:umklapp_config}. In those regions of momentum space, the Ising-nematic form factor is generically nonzero, so the cold-spot suppression is not expected to eliminate the soft-phonon mechanism relevant for current relaxation. Thus the isotropic treatment developed above should still provide a reasonable description of the leading transport scaling, even though the full momentum-resolved phonon softening in the presence of anisotropic form factors remains an interesting problem for future work.

\section{Conclusion}
\label{sec:conclusion}
In this work, we investigated whether softened optical phonons can provide a mechanism for asymptotic low-temperature linear-in-$T$ resistivity. Our analysis separates this question into two logically distinct steps. First, treating the renormalized phonon spectrum as an effective low-energy input, we determined when softened optical phonons can produce linear-in-$T$ single-particle and transport scattering rates. Second, we examined whether such a phonon spectrum can arise microscopically from coupling the phonon to an electronic quantum critical mode.

The general criterion derived in Sec.~\ref{sec:electronphonon} is more restrictive than simple softening of the optical gap. Once Landau damping is included, obtaining asymptotic low-temperature linear-in-$T$ scattering requires not only $\omega_D\to 0$, but also that the renormalized phonon dynamical exponent satisfy $z_p>d$, where $d$ is the spatial dimensionality of the phonon dispersion. Physically, this condition ensures that the phonons which remain in the effective equipartition regime as $T\to 0$ occupy enough momentum-space volume to dominate electron scattering. When this condition fails, the system instead exhibits a sequence of crossover regimes whose temperature dependences are generally stronger than linear. For transport, these results further require that softened phonons can relax momentum efficiently, for example through umklapp processes supplemented by weak disorder.

We then tested this criterion in a concrete model in which an optical phonon couples nonlinearly to a $\vec Q=0$ electronic critical mode. In the clean saddle-point theory with $z_\varphi=3$, the induced phonon softening gives $z_p=2$ in the quasi-2D cases and $z_p=3$ in 3D. As a result, scenario (ii), with quasi-2D electrons and 3D phonons, remains on the unfavorable side of the criterion, while scenarios (i) and (iii) lie only on the marginal boundary. In those marginal cases, the softened phonon enhances low-temperature scattering, but the clean theory does not generically produce a parametrically broad regime of strict asymptotic linear-in-$T$ behavior. Thus, within the leading large-$N$ analysis, the mechanism is at best borderline as an explanation for low-temperature strange-metal transport.

We also examined several effects beyond the simplest clean isotropic treatment. The leading feedback of the softened phonon on the electronic critical mode is marginal in scenarios (i) and (iii), but irrelevant in scenario (ii), and in the marginal cases it is consistent with a tendency toward larger effective $z_\varphi$, which would further reduce $z_p$. Weak disorder modifies the infrared dynamics of the critical mode and can enhance the tendency toward phonon softening, especially in 2D, but it also introduces a low-temperature cutoff to the clean scattering regimes and suppresses the electron-phonon scattering rate in the disorder-dominated regime. Restoring a nontrivial form factor makes the phonon softening anisotropic and calls for a patchwise analysis, although the transport mechanism discussed here is still expected to be controlled primarily by the umklapp-active regions of the Fermi surface.

There are several important limitations of the present work, which also point to natural future directions. First, in analyzing transport we focused on the contribution from softened phonons, and therefore did not include the direct umklapp contribution from the quantum critical boson itself, as studied in Refs.~\cite{PALee2021,PALee2024}. In a realistic system, these two channels should contribute additively to the resistivity, and the dominant mechanism will depend on microscopic details. An interesting direction for future work is therefore to identify models in which the phonon contribution is parametrically enhanced relative to the direct quantum-critical-boson contribution. One possible route is suggested by the  discussion of Sec.~\ref{sec:formfactor}: if the form factor suppresses the coupling of the critical boson near the umklapp hot spots, then the boson-mediated umklapp channel may be weakened, while phonon softening and phonon-assisted transport remain appreciable. Establishing such a regime would require a fully anisotropic treatment of both the critical fluctuations and the phonon sector.

A second direction is to extend the analysis to acoustic phonons. As emphasized in Sec.~\ref{sec:model_softening}, the Goldstone nature of an acoustic mode qualitatively changes the allowed low-energy couplings and can lead to anisotropic softening patterns rather than the more isotropic optical-phonon softening emphasized here. This raises the possibility of direction-selective phonon anomalies near electronic criticality \cite{IPaul2017,VSDeCarvalho2019}, which could be probed experimentally through ultrasound \cite{mason,morse,Pippard,Blount,tsuneto,Allen_US,PAL_sound,ABBhatia1961} or other direction-resolved measurements of lattice dynamics.

\change{It would also be interesting to consider more unconventional phonon degrees of freedom. Chiral phonons carry angular momentum \cite{LZhang2014,HZhu2018} and can possess orbital magnetic moments \cite{DMJuraschek2019}, suggesting possible couplings to spin, magnetic order, or other time-reversal-breaking electronic orders such as loop-current order. Separately, topological mechanical systems can host protected zero modes and extended low-energy manifolds, as in the Kane-Lubensky construction for isostatic lattices \cite{CLKane2014}. Related electron-phonon models can also use vibrational degrees of freedom to generate emergent gauge-field or resonating-valence-bond phases \cite{ZHan2025}. These examples suggest several ways in which coupling itinerant electrons to chiral, topological, or otherwise constrained phonon structures could change the low-energy phonon phase space beyond the isolated-$\Gamma$-point softening analyzed here.}

Finally, the present analysis itself suggests that the phonon and the electronic critical boson should ultimately be treated as a strongly coupled system \cite{YWerman2017,YWerman2017a,ETulipman2020a,ETulipman2021,HGuo2019}. In this work, much of the analysis was organized by first determining the critical boson dynamics and then studying the induced phonon softening, with feedback incorporated only at leading order. The results indicate, however, that this separation is only a first step: in the most interesting regimes, the softened phonon and the critical mode can significantly reshape one another. A more complete theoretical treatment would therefore study the coupled electron-critical-boson-phonon problem on equal footing, for example through a controlled renormalization-group or self-consistent strong-coupling framework.

\section*{Acknowledgements}
We thank Xuepeng Wang for collaboration at early stages of this work and for drawing our attention to a number of relevant experimental references. We thank S. Kivelson, D. Maslov, B. Ramshaw and T. Senthil for insightful discussions. HG is supported by the Bethe-Wilkins-KIC fellowship at Cornell University, and the Kavli Institute at Cornell for Nano Science. DC is funded in part by a NSF CAREER grant (DMR-2237522). 


\begin{appendix}
\numberwithin{equation}{section}
\section{Exactly solvable model}\label{sec:YSYK}

In this appendix, we present the large-$N$ limit of the model described in Eq.~\eqref{eq:action}, which becomes exactly soluble in the $N\to\infty$ limit. We engineer an extension of the Yukawa-SYK model \cite{IEsterlis2021,HGuo2022a,AAPatel2023,ZDShi2023,EEAldape2022}, which generalizes the Yukawa couplings to Gaussian random numbers. The action now reads 
\begin{subequations}
\label{eq:action_largeN}
\beq\label{eq:Le_largeN}
   \calL_e & =& \int \rd\tau \Big[\sum_i \sum_{\vec{k},\sigma} c_{\vec{k},\sigma,i}^\dagger \left(\partial_\tau+\varepsilon_{\vec{k}}-\mu\right)c_{\vec{k},\sigma,i} + \sum_i\sum_{\sigma}\int_\vec{x} \frac{\epsilon_{ij}^{(4)}}{\sqrt{N}}V_\vec{x} c_{\vec{x},\sigma,i}^\dagger c_{\vec{x},\sigma,j} \nn\\
   &&+\sum_l \sum_{\vec{q},\alpha} \varphi_{\alpha,-\vec{q},l}\left(-\partial_\tau^2+v_{\vp}^2\vec{q}_\text{2D}^2+ r\right)\varphi_{\vec{q},\alpha,l} \nn \\
  &&  + \frac{g}{N} \sum_{ijl}\sum_{\vec{k},\vec{q},\sigma,\sigma',\alpha} \epsilon_{ijl}^{(1)} f_{\vec{k},\vec{{q}},\alpha}^{\sigma\sigma'} c_{\vec{k+q/2},\sigma,i}^\dagger c^{\phantom\dagger}_{\vec{k-q/2,j},\sigma'}\varphi_{\alpha,\vec{q},l}\Big]+ ..., \,\\
     \label{eq:Lp_largeN}
  \calL_{p} &=& \int\rd \tau \sum_i\sum_{\q,a}X_{a,-\vec{q},i} \left(\partial_{\tau}^2+c^2\vec{q}^2 + (\omega_0)^2\right) X_{a,\vec{q},i}\,,\\
  \label{eq:Lep_largeN}\calL_{ep} &=& \frac{\lambda}{N}\int\rd \tau \sum_{ijl}\sum_{\vec{k},\vec{q},\sigma,a} \epsilon_{ijl}^{(2)}h^a_{\vec{k},\vec{q}}c^\dagger_{\vec{k+q/2},\sigma,i}c^{\phantom\dagger}_{\vec{k-q/2},\sigma,j} X_{a,\vec{q},l} \\
  &&+ \frac{u}{2N}\int \rd \tau \sum_{ijl} \sum_{\vec{k},\vec{{q}},a,\alpha,\beta} \epsilon_{ijl}^{(3)} L^{a}_{\vec{k},\vec{q},\alpha\beta} \varphi_{\vec{-k+q/2},\beta,i} \varphi_{\vec{k+q/2},\alpha,j} X_{a,\vec{q},l}.\,\nn 
\eeq
\end{subequations} Here we have added an additional flavor index $i,j,l=1,\dots,N$ to the fields, and multiplied the couplings by $\epsilon_{ijl}^{(1,2,3,4)}$. Here, $\epsilon^{(1,2)}_{ijl}$ are complex with the constraint 
$\epsilon^{(1,2)}_{ijl}=\left(\epsilon^{(1,2)}_{jil}\right)^*$,  $\epsilon^{(3)}_{ijl}$ is real with $\epsilon^{(3)}_{ijl}=\epsilon^{(3)}_{jil}$, and $\epsilon_{ij}^{(4)}$ is complex with $\epsilon_{ij}^{(4)}=\left(\epsilon_{ji}^{(4)}\right)^*$. Apart from these constraints, $\epsilon^{1,2,3,4}$ are random numbers drawn from a Gaussian ensemble with zero mean and unit variance. 

In the large-$N$ limit, the saddle point of the above action yields the Migdal-Eliashberg equations for the self-energies. We assume the fermions are $\rm{SU}(2)$-symmetric and focus on the case of a single collective $\varphi$ mode and a single phonon $X$ mode. The self-energies are
\begin{eqnarray}
  \Sigma_e &=&\Sigma_{ee}+\Sigma_{ep}+\Sigma_{d}\,, \\
 \Pi_\varphi &=& \Pi_{\varphi c}+\Pi_{\varphi X}\,, \\
  \Pi_X &=& \Pi_{X c}+\Pi_{X \varphi}\,. 
\end{eqnarray} 
$\Sigma_{ee}$ arises from electron-electron interaction, i.e. the Yukawa coupling to $\varphi$
\begin{equation}\label{}
  \Sigma_{ee}(i\omega,\vec{k})= g^2 \int\frac{\rd \Omega}{2\pi}\frac{\rd^3\vec{q}}{(2\pi)^3} G(i\omega+i\Omega,\vec{k}+\vec{q})D_\varphi(i\Omega,\vec{q})f_{\vec{k}+\vec{q}/2,\vec{q}}^2\,.
\end{equation}
$\Sigma_{ep}$ arises from the deformation-potential coupling to the phonons
\begin{equation}\label{}
  \Sigma_{ep}(i\omega,\vec{k})=\lambda^2 \int \frac{\rd \Omega}{2\pi}\frac{\rd^3\vec{q}}{(2\pi)^3} G(i\omega+i\Omega,\vec{k}+\vec{q})D_X(i\Omega,\vec{q}) h_{\vec{k}+\vec{q}/2,\vec{q}}^2\,.
\end{equation} 
$\Sigma_d$ arises due to the impurity potential 
\begin{equation}\label{}
  \Sigma_d(i\omega)=V^2\int\frac{\rd^3\vec{k}}{(2\pi)^3}G(i\omega,\vec{k})=-\frac{i\Gamma}{2}\sgn\omega\,,
\end{equation} where $\Gamma=2\pi V^2\calN$ is the elastic scattering rate, and $\calN$ is the fermion DOS at the FS.
$\Pi_{\varphi c}$ describes the Landau damping of $\varphi$ due to the FS
\begin{equation}\label{}
  \Pi_{\varphi c}(i\Omega,\vec{q}) = - 2g^2 \int \frac{\rd \omega}{2\pi}\frac{\rd^3\vec{k}}{(2\pi)^3} G(i\omega+i\Omega,\vec{k}+\vec{q})G(i\omega,\vec{k})f_{\vec{k}+\vec{q}/2,\vec{k}}^2\,.
\end{equation}
$\Pi_{\varphi X}$ arises from the coupling between the collective mode and the phonon 
\begin{equation}\label{}
  \Pi_{\varphi X}(i\Omega,\vec{q})=u^2 \int\frac{\rd \omega}{2\pi}\frac{\rd^3 \vec{k}}{(2\pi)^3} D_{\varphi}(i\omega+i\Omega,\vec{k}+\vec{q}) D_X(i\omega,\vec{k}) L_{\vec{k}+\vec{q}/2,\vec{q}}^2\,.
\end{equation}
$\Pi_{X c}$ describes the Landau damping of the phonon $X$ due to the FS
\begin{equation}\label{}
  \Pi_{X c}(i\Omega,\vec{q}) = - 2\lambda^2 \int \frac{\rd \omega}{2\pi}\frac{\rd^3\vec{k}}{(2\pi)^3} G(i\omega+i\Omega,\vec{k}+\vec{q})G(i\omega,\vec{k})h_{\vec{k}+\vec{q}/2,\vec{k}}^2\,.
\end{equation}
$\Pi_{X \varphi}$ describes the renormalization of the phonon dispersion due to the collective mode $\varphi$
\begin{equation}\label{}
  \Pi_{X \varphi}(i\Omega,\vec{q})=\frac{u^2}{2} \int\frac{\rd \omega}{2\pi}\frac{\rd ^3\vec{k}}{(2\pi)^3} D_\varphi(i\omega+i\Omega,\vec{k}+\vec{q}) D_\varphi(i\omega,\vec{k}) L_{\vec{k}+\vec{q}/2,\vec{q}}^2\,.
\end{equation}

In the equations above, the self-energy components are written using the full Green's functions $G,D_\varphi,D_X$ of the fields, which are related to the self-energies via Schwinger-Dyson equations 
\begin{eqnarray}\label{}
  G(i\omega,\vec{k}) &=& \frac{1}{i\omega-\xi_{\vec{k}}-\Sigma_{e}(i\omega,\vec{k})}\,, \\
  D_\varphi(i\Omega,\vec{q}) &=& \frac{1}{\Omega^2+v_\varphi^2 \vec{q_\text{2D}^2}+r-\Pi_\varphi(i\Omega,\vec{q})} \,, \\
  D_X(i\Omega,\vec{q}) &=& \frac{1}{\Omega^2+c^2\vec{q}^2+\omega_0^2-\Pi_{X}(i\Omega,\vec{q})} \,.
\end{eqnarray} Here $\xi_\vec{k}=\varepsilon_{k}-\mu$. In the above derivation, $\int \rd \omega/(2\pi)$ can denote either continuous integration at $T=0$ or discrete Matsubara sums at $T\neq 0$. We have assumed the dispersions of $c$ and $\varphi$ are quasi-2D, and the generalization to 3D case is straightforward.

\section{Computation of electron-phonon scattering rate}\label{sec:funcI}

In this appendix, we derive the scaling forms for the electron-phonon scattering rate $\Gamma_\text{ep}$ that were summarized in Sec.~\ref{sec:electronphonon}, specifically in Fig.~\ref{fig:selfenergy1}, and Table.~\ref{tab:selfenergy1}. Our starting point is Eq.~\eqref{eq:Gammaep1} of the main text,
\begin{equation}\label{eq:Gammaep_app}
  \Gamma_\text{ep}(\vec{k})=-\lambda^2 \int\frac{\rd^3 \vec{q}}{(2\pi)^3} \int \frac{\rd z}{2\pi} \frac{1}{\sinh \beta z} A_F(z,\vec{k}+\vec{q}) A_X(z,\vec{q})\,.
\end{equation}
The strategy is the following. First, we reduce the scattering rate to a momentum integral over a single function
$I(\beta,\omega_{\vec q},\eta_{\vec q})$, which measures the contribution of an individual phonon mode. Second, we determine the asymptotic forms of $I$ in the overdamped and underdamped regimes. Finally, we translate those asymptotics into momentum-space crossover scales and evaluate the resulting $q$-integrals in the three geometries considered in the main text.

\subsection{2D Fermi surface + 2D phonon}

We begin with the case of a quasi-2D Fermi surface coupled to a quasi-2D softened optical phonon,
\begin{equation}
  \omega_\vec{q}^2=C^2\vn{q}^{z_p-1}+\omega_D^2\,.
\end{equation}
Since the integrand in Eq.~\eqref{eq:Gammaep_app} is independent of $q_z$, the $q_z$ integral simply produces $\int \rd q_z/(2\pi)=1/a_z$.

To reduce the remaining 2D momentum integral, we introduce an auxiliary momentum $\vec{k'}$ through
\[
1=\int \rd^2\vec{k'}\,\delta(\vec{k'}-\vec{k}-\vec{q})\,.
\]
Here $\vec{k}$, $\vec{k'}$, and $\vec{q}$ are all understood as in-plane momenta. After rewriting the $\vec{k'}$ and $\vec{q}$ integrals in polar coordinates, the angular integrals can be done explicitly:
\[
  \int \rd\theta_q \rd \theta_{k'}\delta(\vec{k'}-\vec{q}-\vec{k})= \frac{4}{\sqrt{(\vn{k}+\vn{k'})^2-\vn{q}^2}\sqrt{\vn{q}^2-(\vn{k}-\vn{k'})^2}}\,.
\]
Near the Fermi surface, $\vn{k}\sim \vn{k'}\sim k_F$, so the first square root may be approximated by $2k_F$. The second is approximated by $\vn{q}$ because
$\vn{k}-\vn{k'}\sim \xi_k/v_F\sim \Sigma_e(i\omega)/v_F\ll \vn{q}$ in the scaling regime of interest; this can be checked \emph{a posteriori} \cite{HGuo2022a,HGuo2024d}. Thus
\begin{equation}\label{eq:Gammaep_B3}
  \Gamma_\text{ep}(\vec{k})=-\frac{\lambda^2}{a_z} \int\frac{\vn{k'}\rd\vn{k'}}{2\pi}\frac{\vn{q}\rd\vn{q}}{2\pi} \int \frac{\rd z}{2\pi}\frac{1}{\sinh \beta z} \frac{2}{k_F \vn{q}}A_F(z,\vn{k'})A_X(z,\vn{q})\,.
\end{equation}

The $\vn{k'}$ integral is now straightforward, since $A_F(z,\vn{k'})$ is sharply peaked near $k_F$ \cite{REPrange1964,YBKim1995a,CPNave2007,PALee2021,HGuo2022a}:
\begin{equation}
  \frac{1}{a_z}\int \frac{\vn{k'}\rd\vn{k'}}{(2\pi)^2} A_F(z,\vn{k'})=\calN_\text{2D}\,,
\end{equation}
where $\calN_\text{2D}=k_F/(2\pi v_F a_z)$ is the density of states per spin. Therefore
\begin{equation}\label{eq:Gammaep_app2}
  \Gamma_\text{ep}=\frac{2\lambda^2\calN_\text{2D}}{k_F} \int_0^{\Lambda} \rd\vn{q}\, I(\beta,\omega_\vec{q},\eta_\vec{q})\,,
\end{equation}
with
\begin{equation}\label{eq:int_I}
  I(\beta,\omega_\vec{q},\eta_\vec{q})\equiv -\int \frac{\rd z}{2\pi}\frac{1}{\sinh \beta z} A_X(z,\vn{q})\,,
\end{equation}
and
\begin{equation}
  -A_X(z,\vn{q})=\frac{2 z \gamma \eta_\vec{q}}{z^2 \gamma^2+\eta_\vec{q}^2 (z^2-\omega_\vec{q}^2)^2}\,.
\end{equation}
For the 2D Fermi surface, $\gamma=2\lambda^2\calN_\text{2D}$ and $\eta_\vec{q}=v_F\vn{q}$.

The exact expression for $I$ can be obtained by decomposing $A_X$ into partial fractions and evaluating the resulting contour integrals:
\begin{equation}
  I(\beta,\omega_\vec{q},\eta_\vec{q})=\frac{i}{2\pi\sqrt{4\omega_\vec{q}^2-\gamma^2/\eta_\vec{q}^2}}\left[\calF\left(\beta z_1\right)-\calF\left(\beta z_2\right)\right]\,,
\end{equation}
where
\begin{equation}
\begin{split}
  \calF(a)&=\frac{2\pi \sgn \Re a}{\sin a}+\psi\left(\frac{a}{2\pi}+\frac{1}{2}\right)+\psi\left(\frac{-a}{2\pi}+\frac{1}{2}\right)\\
  &\quad-\psi\left(\frac{a}{2\pi}\right)-\psi\left(\frac{-a}{2\pi}\right)\,,
\end{split}
\end{equation}
and
\begin{equation}
  z_{1,2}=\frac{\gamma}{2\eta_\vec{q}}\pm i\sqrt{\omega_\vec{q}^2-\frac{\gamma^2}{4 \eta_{\vec{q}}^2}}\,.
\end{equation}
The exact form is less important than its asymptotics. For our purposes, $I$ is simply the scattering rate associated with a phonon mode at momentum $\vec q$, and it interpolates between a linear-in-$T$ form and a quadratic-in-$T$ form depending on the damping regime.

When the phonon is overdamped, $\omega_\vec{q}<\gamma/(2\eta_\vec{q})$, the poles become real. Writing
\[
z_{1}\approx \frac{\eta_\vec{q}\omega_\vec{q}^2}{\gamma},\qquad
z_2\approx \frac{\gamma}{\eta_\vec{q}},
\]
the asymptotic forms are
\begin{subequations}\label{eq:Is}
\begin{numcases}{I(\beta,\omega_\vec{q},\eta_\vec{q})\to}
    \frac{\pi \gamma}{2\eta_\vec{q}\omega_\vec{q}^4 \beta^2}\,, &  $\omega_\vec{q}<\frac{\gamma}{2\eta_\vec{q}}\,,1/\beta\ll z_1\,,$ \label{eq:I1} \\ 
    \frac{1}{\beta \omega_\vec{q}^2}\,, & $\omega_\vec{q}<\frac{\gamma}{2\eta_\vec{q}}\,, z_1\ll 1/\beta \ll z_2\,,$ \label{eq:I2}\\
    \frac{1}{\beta \omega_\vec{q}^2}\,, & $\omega_\vec{q}<\frac{\gamma}{2\eta_\vec{q}}\,, z_2\ll 1/\beta \,.$ \label{eq:I3}
\end{numcases}
\end{subequations}
The second and third lines have the same leading $T$-dependence but differ at subleading order.

When the phonon is underdamped, $\omega_\vec{q}>\gamma/(2\eta_\vec{q})$, the poles are complex with $|z_{1,2}|=\omega_\vec{q}$, and
\begin{subequations}
\begin{numcases}{I(\beta,\omega_\vec{q},\eta_\vec{q})\to}
 \quad\frac{\pi \gamma}{2\eta_\vec{q}\omega_\vec{q}^4 \beta^2}+ & $\omega_\vec{q}>\frac{\gamma}{2\eta_\vec{q}}\,,1/\beta\ll\omega_\vec{q}\,,$  \\
 \frac{2e^{-\beta\sqrt{\omega_\vec{q}^2-\gamma^2/(4\eta_\vec{q}^2)}}}{\sqrt{\omega_\vec{q}^2-\gamma^2/(4\eta_\vec{q}^2)}}\cos\frac{\beta\gamma}{2\eta_\vec{q} }, & \nonumber   \\
 \frac{1}{\beta \omega_\vec{q}^2}, & $ \omega_\vec{q}>\frac{\gamma}{2\eta_\vec{q}}\,, \omega_\vec{q}\ll 1/\beta\,.$ \label{eq:I5}
\end{numcases}
\end{subequations}
The exponentially activated piece is negligible for the scaling analysis below.

It is more transparent to translate the asymptotic forms of $I(\beta,\omega_{\vec q},\eta_{\vec q})$ into momentum space. At fixed temperature, the contribution of a given phonon mode depends on where its momentum lies relative to several crossover scales. These scales separate regimes in which the phonon is gap dominated or dispersion dominated, overdamped or underdamped, and effectively classical or quantum mechanical.

The first scale is
\begin{equation}
  q_\text{gap}\equiv \left(\frac{\omega_D}{C}\right)^{\frac{2}{z_p-1}}\,,
\end{equation}
defined by the condition
\[
C^2 q_\text{gap}^{\,z_p-1}\sim \omega_D^2.
\]
For $q\ll q_\text{gap}$, the phonon frequency is controlled mainly by the optical gap, $\omega_{\vec q}\approx \omega_D$. For $q\gg q_\text{gap}$, the softened dispersion dominates and $\omega_{\vec q}$ acquires its critical momentum dependence.

The second scale is the overdamped--underdamped crossover momentum,
\begin{equation}
  q_\text{od}:~ \frac{\gamma}{2 v_F q_\text{od}}=\sqrt{\omega_D^2+C^2 q_\text{od}^{\,z_p-1}}\,.
\end{equation}
This is obtained by comparing the damping rate $\gamma/(2\eta_{\vec q})=\gamma/(2v_F q)$ with the phonon frequency $\omega_{\vec q}$. For $q\ll q_\text{od}$, the phonon is overdamped; for $q\gg q_\text{od}$, it is underdamped.

Inside the overdamped sector, there is a further scale
\begin{equation}
  q_1:~\frac{\eta_{\vec q}\omega_{\vec q}^2}{\gamma }=T\,.
\end{equation}
This is the momentum at which the smaller pole $z_1\sim \eta_{\vec q}\omega_{\vec q}^2/\gamma$ crosses the thermal scale $T$. For $q\lesssim q_1$, the corresponding phonon modes contribute with the classical form
\[
I(\vec q)\sim \frac{T}{\omega_{\vec q}^2},
\]
whereas for $q\gtrsim q_1$ the contribution crosses over to the Fermi-liquid-like form
\[
I(\vec q)\sim \frac{\gamma T^2}{\eta_{\vec q}\omega_{\vec q}^4}.
\]

Finally, in the underdamped sector one can define the thermal momentum
\begin{equation}
  q_T:~ \omega_{\vec q}=T\,.
\end{equation}
This scale exists only when $T>\omega_D$. For $q\ll q_T$, the underdamped phonons are thermally occupied and again contribute as
\[
I(\vec q)\sim \frac{T}{\omega_{\vec q}^2},
\]
whereas for $q\gg q_T$ they are effectively frozen out and contribute only through the subleading $T^2$ tail.

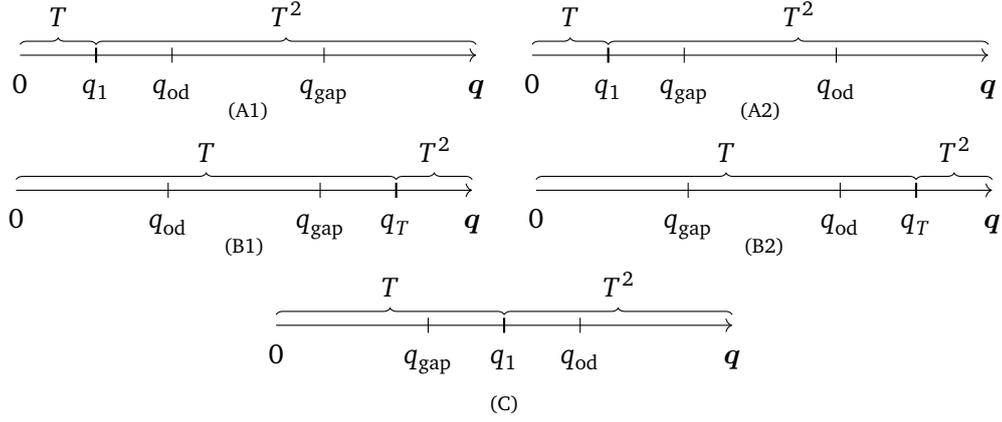
\begin{figure}
\centering
\begin{subfigure}[t]{0.4\textwidth}
\begin{tikzpicture}
  \draw[->] (0,0)--(6,0);
  \node[position label] (qf) at  (6,0) {$\vec{q}$};
  \node[position label] (qi) at  (0,0)  {0};
  \draw (2,0.1)--(2,-0.1);
  \draw (4,0.1)--(4,-0.1);
  \node[position label] at (2,0) {$q_\text{od}$};
  \node[position label] at (4,0) {$q_\text{gap}$};
  \draw[thick] (1,0.1)--(1,-0.1);
  \node[position label] (qs) at (1,0) {$q_1$};
  \draw[brace] (qi.north)--(qs.north) node[pos=0.5, position label,above=5pt] {$T$};
  \draw[brace] (qs.north)--(qf.north) node[pos=0.5, position label,above=5pt] {$T^2$};
  \node[below] at (3,-0.5) {\scriptsize (A1)};
\end{tikzpicture}
\end{subfigure}\qquad 
\begin{subfigure}[t]{0.4\textwidth}
\begin{tikzpicture}
  \draw[->] (0,0)--(6,0);
  \node[position label] (qf) at  (6,0) {$\vec{q}$};
  \node[position label] (qi) at  (0,0)  {0};
  \draw (2,0.1)--(2,-0.1);
  \draw (4,0.1)--(4,-0.1);
  \node[position label] at (2,0) {$q_\text{gap}$};
  \node[position label] at (4,0) {$q_\text{od}$};
  \draw[thick] (1,0.1)--(1,-0.1);
  \node[position label] (qs) at (1,0) {$q_1$};
  \draw[brace] (qi.north)--(qs.north) node[pos=0.5, position label,above=5pt] {$T$};
  \draw[brace] (qs.north)--(qf.north) node[pos=0.5, position label,above=5pt] {$T^2$};
  \node[below] at (3,-0.5) {\scriptsize (A2)};
\end{tikzpicture}
\end{subfigure}

\begin{subfigure}[t]{0.4\textwidth}
\begin{tikzpicture}
  \draw[->] (0,0)--(6,0);
  \node[position label] (qf) at  (6,0) {$\vec{q}$};
  \node[position label] (qi) at  (0,0)  {0};
  \draw (2,0.1)--(2,-0.1);
  \draw (4,0.1)--(4,-0.1);
  \node[position label] at (2,0) {$q_\text{od}$};
  \node[position label] at (4,0) {$q_\text{gap}$};
  \draw[thick] (5,0.1)--(5,-0.1);
  \node[position label] (qs) at (5,0) {$q_T$};
  \draw[brace] (qi.north)--(qs.north) node[pos=0.5, position label,above=5pt] {$T$};
  \draw[brace] (qs.north)--(qf.north) node[pos=0.5, position label,above=5pt] {$T^2$};
  \node[below] at (3,-0.5) {\scriptsize (B1)};
\end{tikzpicture}
\end{subfigure}
\qquad
\begin{subfigure}[t]{0.4\textwidth}
\begin{tikzpicture}
  \draw[->] (0,0)--(6,0);
  \node[position label] (qf) at  (6,0) {$\vec{q}$};
  \node[position label] (qi) at  (0,0)  {0};
  \draw (2,0.1)--(2,-0.1);
  \draw (4,0.1)--(4,-0.1);
  \node[position label] at (2,0) {$q_\text{gap}$};
  \node[position label] at (4,0) {$q_\text{od}$};
  \draw[thick] (5,0.1)--(5,-0.1);
  \node[position label] (qs) at (5,0) {$q_T$};
  \draw[brace] (qi.north)--(qs.north) node[pos=0.5, position label,above=5pt] {$T$};
  \draw[brace] (qs.north)--(qf.north) node[pos=0.5, position label,above=5pt] {$T^2$};
  \node[below] at (3,-0.5) {\scriptsize (B2)};
\end{tikzpicture}
\end{subfigure}

\begin{subfigure}[t]{0.4\textwidth}
\begin{tikzpicture}
  \draw[->] (0,0)--(6,0);
  \node[position label] (qf) at  (6,0) {$\vec{q}$};
  \node[position label] (qi) at  (0,0)  {0};
  \draw (2,0.1)--(2,-0.1);
  \draw (4,0.1)--(4,-0.1);
  \node[position label] at (2,0) {$q_\text{gap}$};
  \node[position label] at (4,0) {$q_\text{od}$};
  \draw[thick] (3,0.1)--(3,-0.1);
  \node[position label] (qs) at (3,0) {$q_1$};
  \draw[brace] (qi.north)--(qs.north) node[pos=0.5, position label,above=5pt] {$T$};
  \draw[brace] (qs.north)--(qf.north) node[pos=0.5, position label,above=5pt] {$T^2$};
  \node[below] at (3,-0.8) {\scriptsize (C)};
\end{tikzpicture}
\end{subfigure}
\caption{Momentum-space structure of $I(\vec q)$ in the different crossover regimes. The label $T$ denotes $I(\vec q)\sim T/\omega_\vec{q}^2$, while $T^2$ denotes $I(\vec q)\sim \gamma T^2/(\eta_\vec{q}\omega_\vec{q}^4)$. Panels (A1) and (A2) correspond to the Fermi-liquid regime with $\omega_D\gtrless \omega_*$, panels (B1) and (B2) to the generalized Bloch-Gr\"uneisen regime with $\omega_D\gtrless \omega_*$, and panel (C) to the electron-phonon critical regime.}
\label{fig:I}
\end{figure}

Collecting these results, $I(\vec q)$ is built from two basic behaviors:
\[
I(\vec q)\sim \frac{T}{\omega_{\vec q}^2}
\qquad\text{or}\qquad
I(\vec q)\sim \frac{\gamma T^2}{\eta_{\vec q}\omega_{\vec q}^4},
\]
and the hierarchy among $q_\text{gap}$, $q_\text{od}$, $q_1$, and $q_T$ determines which of these behaviors dominates the momentum integral for $\Gamma_\text{ep}$. The different possible orderings are summarized schematically in Fig.~\ref{fig:I}.

We now address the regimes introduced in Fig.~\ref{fig:selfenergy1}, which corresponds to different hierarchies of the momentum scales introduced above. We divide the regimes (A), (B) into subregimes (A1,A2) and (B1,B2) respectively, depending on whether $\omega_D\gg \omega_*$ or $\omega_D\ll \omega_*$.

\subsubsection{FL Regime (A)}

 We first consider the FL regime (A). The first subregime is defined  with $q_\text{gap}\gg q_\text{od}$, and we label this regime (A1). This condition can be translated to $\omega_D\gg \omega_*$, where $\omega_*\sim C^{2/(z_p+1)}(\gamma/v_F)^{(z_p-1)/(z_p+1)}$. In this regime, we have 
$$
q_\text{od} \sim \frac{\gamma}{v_F \omega_D}\,,\quad q_1\sim q_\text{od}\frac{T}{\omega_D}\,.
$$

Therefore, when $T\ll \omega_D$,  we have $q_1\ll q_\text{od}$, and the behavior of function $I(\vec{q})$ is summarized in Fig.~\ref{fig:I} (A1). Therefore $\Gamma_\text{ep}$ is estimated to be 
\begin{equation}\label{}
\begin{split}
  \Gamma_\text{ep}&\sim \frac{\gamma}{k_F}\left[T\int_0^{q_1} \frac{\rd q}{\omega_q^2}+\frac{\gamma T^2}{v_F} \int_{q_1}^\Lambda \frac{\rd q}{q \omega_q^4}\right] \,.
\end{split}
\end{equation} We have dropped the $\calO(1)$ numerical factors. The first term yields 
\begin{equation}\label{}
  \frac{\gamma^2 T^2}{k_F v_F \omega_D^4}\,,  \nonumber
\end{equation} where we have used $\gamma=2\calN_\text{2D}\lambda^2$.  The second term requires a bit more care due to the log-divergence in the lower limit. Inserting the dispersion $\omega_\vec{q}^2$ into the integral, we obtain the logarithmic term to be 
\begin{equation}\label{}
  \frac{\gamma^2 T^2}{k_F v_F \omega_D^4} \left.\ln\frac{C^2 q^{z_p-1}}{C^2 q^{z_p-1}+\omega_D^2}\right|^{q=\Lambda}_{q=q_1}\,. \nonumber
\end{equation} The $q=\Lambda$ term only contributes a $\calO(1)$ factor, and the $q=q_1$ term leads to 
\begin{equation}\label{}
  \frac{\gamma^2 T^2}{k_F v_F \omega_D^4} \ln\frac{q_\text{gap}}{q_1}\sim \frac{\gamma^2 T^2}{k_F v_F \omega_D^4} \ln \frac{T_\text{FL}}{T}\,, \nonumber
\end{equation} where 
\begin{equation}\label{}
  T_\text{FL}\sim \frac{v_F \omega_D^2 q_\text{gap}}{\gamma}\,.
\end{equation}

Assembling everything, the leading dependence of $\Gamma_\text{ep}$ is 
\begin{equation}\label{eq:B18}
  \Gamma_\text{ep}\sim \frac{\gamma^2 T^2}{k_F v_F \omega_D^4} \ln\frac{T_\text{FL}}{T}\,,
\end{equation} which is Eq.\eqref{eq:Gammaep_A_2D} of main text. The crossover out of the FL regime happens when $q_1=q_\text{od}$, i.e. $T\sim \omega_D$. We also note that  $\omega_D\gg \omega_*$ implies $T_\text{FL}\gg \omega_D$, so the argument of the logarithmic is always large.

We discuss the other half of the FL regime (A2), which is defined by the condition $q_\text{gap}\ll q_\text{od}$, or equivalently $\omega_D\ll\omega_*$. The hierarchy of momentum scales is shown in Fig.~\ref{fig:I} (A2). In this regime, $\Gamma_\text{ep}$ is also described by Eq.\eqref{eq:B18}. The difference of this regime with (A1) is the boundary of the crossover. Here, the boundary is set by the condition $q_1=q_\text{gap}$, which is equivalent to $T\sim T_\text{Fl}$.

To summarize, the boundary of the FL regime is set by 
\begin{equation}\label{}
  T \ll \min(T_\text{FL},\omega_D)\,,
\end{equation} and $\Gamma_\text{ep}$ is described by Eq.\eqref{eq:B18}.

\subsubsection{Generalized Bloch-Gr\"{u}neisen Regime (B)}

We now turn to the Generalized Bloch-Gr\"{u}neisen Regime (B), which is also refined to two subregimes depending on $q_\text{gap}\gg q_\text{od}$ (B1) or $q_\text{gap}\ll q_\text{od}$ (B2). We discuss (B1) first. 

When we start from the regime (A1) and let $T$ grow beyond $\omega_D$ so that $T\gg \omega_D$,  the scale $q_1$ no longer makes sense, and the crossover momentum scale for $I(\vec{q})$ becomes $q_T$, which by definition satisfies $q_T\gg q_\text{gap}$. Therefore, the system now crosses over to the Generalized Bloch-Gr\"{u}neisen regime (B1), where the main contribution to electron-phonon scattering is from thermally activated phonons. 

The behavior of $I(\vec{q})$ is shown in Fig.~\ref{fig:I} (B1), and the crossover scale $q_T$ is 
$$
q_T \sim \left(\frac{T}{C}\right)^{\frac{2}{z_p-1}} \gg q_\text{gap}\,.
$$
Therefore 
\begin{equation}\label{eq:B19}
\begin{split}
  \Gamma_\text{ep}&\sim \frac{\gamma}{k_F}\left[T\int_0^{q_T} \frac{\rd q}{\omega_q^2}+\frac{\gamma T^2}{v_F} \int_{q_T}^\Lambda \frac{\rd q}{q \omega_q^4}\right] \,.
\end{split}
\end{equation} 
The dominant contribution arises from the first term, which is 
\begin{equation}\label{eq:B20}
  \frac{\gamma T}{k_F}\int_0^{q_T} \frac{\rd q}{\omega_D^2+C^2 q^{z_p-1}}=\left.\frac{\gamma T}{k_F} \frac{1}{(2-z_p)q^{z_p-2}} \pFq{2}{1}{1,\frac{z_p-2}{z_p-1}}{2+\frac{1}{1-z_p}}{-\frac{\omega_D^2}{C^2 q^{z_p-1}}}\right|_{q=0}^{q=q_T}\,.
\end{equation} Here ${}_2 F_1$ is the hypergeometric function. 
The limiting behavior of Eq.\eqref{eq:B20} can be obtained by simple power counting. When $z_p<2$, the integral converges when $\omega_D=0$, yielding 
\begin{equation}\label{eq:B22}
  \Gamma_\text{ep}^{z_p<2}\sim \frac{\gamma T}{k_F} \frac{q_T^{2-z_p}}{C^2}\sim \frac{\gamma T}{k_F C^2} \left(\frac{T}{C}\right)^{\frac{2(2-z_p)}{z_p-1}}\,,
\end{equation} and the correction due to $\omega_D$ vanishes when $\omega_D\to 0$. This is Eq.\eqref{eq:Gammaep_B1} of the main text. 

When $z_p=2$, Eq.\eqref{eq:B20} is logarithmic, yielding 
\begin{equation}\label{eq:B23}
  \Gamma_\text{ep}^{z_p=2}\sim \frac{\gamma T}{k_F C^2} \ln\frac{T}{\omega_D}\,,
\end{equation} which is Eq.\eqref{eq:Gammaep_B2} of the main text.

When $z_p>2$, Eq.\eqref{eq:B20} diverges when $\omega_D\to 0$. The main contribution arises from $0<q<q_\text{gap}$, and we obtain 
\begin{equation}\label{eq:B24}
  \Gamma_\text{ep}^{z_p>2}\sim \frac{\gamma T}{k_F} \frac{q_\text{gap}}{\omega_D^2}\,,
\end{equation} which is Eq.\eqref{eq:Gammaep_E} of the main text, labelled as regime (E). The upper boundary of the regime is set by the condition that $q_T$ should not exceed the UV cutoff $\Lambda$, or equivalently 
\begin{equation}\label{}
  T\ll \sqrt{\omega_D^2+C^2\Lambda^{z_p-1}}\,.
\end{equation}

We now turn to the other half (B2) of the regime, which is defined by the condition $q_\text{gap}\ll q_\text{od}$ or $\omega_D\ll \omega_*$. The hierarchy of momentum scales are shown in Fig.~\ref{fig:I} (B2). The difference between (B2) and (B1) is again the boundary. The lower boundary is now set by $q_T\gg q_\text{od}$, or equivalently $T\gg \omega_*$.

To summarize, in the generalized Bloch Gr\"uneisen regime $\Gamma_\text{ep}$ is given by Eqs.\eqref{eq:B22}-\eqref{eq:B23}, and the boundary of the regime is given by 
\begin{equation}\label{eq:B26}
  \max(\omega_D,\omega_*)\ll T \ll \sqrt{\omega_D^2+C^2 \Lambda^{z_p-1}}\,.
\end{equation}

Finally, we check that the second term of Eq.\eqref{eq:B19} is indeed subleading. We evaluate the integral with the explicit dispersion relation, and expand the result in the limit of $\omega_D^2/(C^2 q_T^{z_p-1})\to 0$. Here, it is also important to keep track of the other regular terms in addition to the log, using 
\begin{equation}\label{}
  \nonumber
  \int \frac{\rd q}{q} \frac{1}{\left(\omega_D^2+C^2 q^{z-1}\right)^2}=\frac{1}{\omega_D^4(z-1)}\left[\frac{\omega_D^2}{ q^{z-1}+\omega_D^2}+\ln\frac{C^2 q^{z-1}}{C^2 q^{z-1}+\omega_D^2}\right]\,,
\end{equation} and we obtain the contribution to $\Gamma_\text{ep}$ to be 
\begin{equation}\label{}
  \frac{\gamma^2}{k_F v_F T^2}\,,
\end{equation} this is much smaller than the first term, because the only relevant energy scales here are $\omega_D,T,\omega_*$, and the Generalized Bloch-G\"{u}neisen regime is defined by $T\gg \omega_D,\omega_*$. 

\subsubsection{Electron-phonon critical regime (C)}

This regime sits in between the FL regime (A2) and the generalized Bloch-Gr\"{u}neisen regime (B2), and the hierarchy of momentum scales is shown in Fig.~\ref{fig:I} (C). The crossover momentum scale for $I(\vec{q})$ is set by 
\begin{equation}\label{}
  q_1\sim \left(\frac{\gamma T}{v_F C^2}\right)^{\frac{1}{z_p}}\,.
\end{equation} The electron-phonon scattering rate $\Gamma_\text{ep}$ is estimated by 
\begin{equation}\label{eq:B29}
\begin{split}
  \Gamma_\text{ep}&\sim \frac{\gamma}{k_F}\left[T\int_0^{q_1} \frac{\rd q}{\omega_q^2}+\frac{\gamma T^2}{v_F} \int_{q_1}^\Lambda \frac{\rd q}{q \omega_q^4}\right] \,.
\end{split}
\end{equation} The evaluation of the above integral is similar to the analysis of regime (B), except that the integration limit is different. The results are 
\begin{equation}\label{}
  \Gamma_\text{ep}^{z_p<2}\sim \frac{\gamma T}{k_F} \frac{q_1^{2-z_p}}{C^2} \sim \frac{v_F}{k_F} \left(\frac{\gamma T}{v_F C^2}\right)^{\frac{2}{z_p}}\,,
\end{equation}
\begin{equation}\label{}
  \Gamma_\text{ep}^{z_p=2}\sim \frac{\gamma T}{k_F C^2} \ln \frac{C^2 \gamma T}{v_F\omega_D^4}\,,
\end{equation} and 
\begin{equation}\label{}
  \Gamma_\text{ep}^{z_p>2}\sim \frac{\gamma T}{k_F} \frac{q_\text{gap}}{\omega_D^2}\,.
\end{equation} We also note that when $0<z_p<2$, both terms in Eq.\eqref{eq:B29} contribute with the same parametric dependence. 

The boundary of the regime is set by $q_\text{gap} \ll q_1\ll q_\text{od}$, which is equivalent to 
\begin{equation}\label{}
  T_\text{FL}\ll T \ll \omega_*\,.
\end{equation}

\subsubsection{Equipartition regime (D) and (E)}

As we further increase the temperature $T$ from regime (B), the scale $q_T$ will eventually reach the UV cutoff $\Lambda\sim 2k_F$, and the system enters the equipartition regime. The estimate for $\Gamma_\text{ep}$ can be obtained by setting $q_T=\Lambda$ in Eq.\eqref{eq:B20}, yielding 
\begin{equation}\label{}
  \Gamma_\text{ep}^{z_p<2} \sim \frac{\gamma T}{k_F} \frac{\Lambda^{2-z_p}}{C^2}\,,\quad \omega_D^2 \ll C^2\Lambda^{z_p-1}
\end{equation} 
\begin{equation}\label{}
  \Gamma_\text{ep}^{z_p=2} \sim \frac{\gamma T}{k_F C^2} \ln\frac{C \Lambda^{\frac{1}{2 }}}{\omega_D}\,,\quad \omega_D^2 \ll C^2\Lambda^{}
\end{equation} and 
\begin{equation}\label{}
  \Gamma_\text{ep}^{z_p>2}\sim \frac{\gamma T}{k_F}\frac{q_\text{gap}}{\omega_D^2}\,, \quad \omega_D^2 \ll C^2\Lambda^{z_p-1} \,.
\end{equation} The results above assume $\omega_D\ll C \Lambda^{(z_p-1)/2}$. In the opposite limit $\omega_D\gg C \Lambda^{(z_p-1)/2}$, we obtain 
\begin{equation}\label{}
  \Gamma_\text{ep}\sim \frac{\gamma T}{k_F}\frac{\Lambda}{\omega_D^2}\,.
\end{equation}

When $z_p\leq 2$, the boundary between the equipartition regime (D) and the Generalized Bloch-Gr\"{u}neisen regime (B) is determined by the condition that $T$ is much larger than the top of the phonon band, i.e. 
\begin{equation}\label{}
  T\gg \sqrt{\omega_D^2+C^2\Lambda^{z_p-1}}\,.
\end{equation}

When $z_p>2$, the equipartition regime is enlarged and merges regimes (B,C,D), so the boundary is now 
\begin{equation}\label{}
  T\gg \max(T_\text{FL},\omega_D)\,.
\end{equation}

\subsection{2D Fermi surface + 3D phonon}

Now, we consider the case 2D FS and 3D phonon. The electronic excitations are still 2D, but the phonon has dispersion in the $z$-direction, given by 
\begin{equation}\label{}
  \omega_\vec{q}^2=C^2\vn{q_\text{2D}}^{z_p-1}+c_z^2 q_z^2+\omega_D^2\,.
\end{equation} Since the fermion dispersion is still 2D, it can be handled with the same manipulations as in the previous subsection, leading to 
\begin{equation}\label{eq:B41}
  \Gamma_\text{ep}=\frac{2\lambda^2\calN_\text{2D}}{k_F} \int_{0}^{\Lambda_z} \frac{\rd q_z}{\Lambda_z} \int_0^\Lambda \rd\vn{q} I(1/T,\omega_\vec{q},\eta_\vec{q})\,.
\end{equation}  Here the z-direction momentum cutoff is $\Lambda_z=\pi/a_z$, where $a_z$ is the lattice constant in the $z$-direction. 

When z-dispersion is weak, $c_z \Lambda_z\ll \omega_D$, we obtain $\Gamma_\text{ep}(\omega_D)\approx \bar{\Gamma}_\text{ep} (\omega_D)$, so the result is similar to the case of 2D phonon. 

In the opposite limit of strong $z$-dispersion, i.e. $c_z \Lambda_z\gg \omega_D$, the behavior is qualitatively different, which we now discuss.

\subsubsection{FL regime (A)}

The result for $\Gamma_\text{ep}$ in this regime can be constructed from the result of the previous section. Let's use $\bar{\Gamma}_\text{ep}(\omega_D)$ to denote the result for $\Gamma_\text{ep}$ in the previous section as a function of $\omega_D$, then Eq.\eqref{eq:B41} can be rewritten as 
\begin{equation}\label{eq:B42}
  \Gamma_\text{ep}(\omega_D)= \frac{1}{c_z \Lambda_z}\int_{\omega_D}^{\sqrt{\omega_D^2+c_z^2 \Lambda_z^2}}  \frac{\omega\rd \omega}{\sqrt{\omega^2-\omega_D^2}}\bar{\Gamma}_\text{ep}(\omega)\approx \frac{1}{c_z \Lambda_z} \int^{c_z \Lambda_z}_{\omega_D} \rd \omega \bar{\Gamma}_\text{ep}(\omega)\,.
\end{equation}

The whole integral \eqref{eq:B42} stays inside regime (A) of the pure 2D case, so we simply integrate Eq.\eqref{eq:B18}, and obtain 
\begin{equation}\label{}
  \Gamma_\text{ep} \sim \frac{\gamma^2}{\omega_D^4} \frac{\omega_D}{c_z \Lambda_z} \frac{T^2}{k_F v_F} \ln\frac{T_\text{FL}}{T}\,.
\end{equation} 

\subsubsection{Generalized Bloch-Gr\"uneisen regime (B)}

In this regime, the trick introduced in Eq.\eqref{eq:B42} is not convenient to use because the previous IR divergent integrals will become convergent after integrating over $\omega$, and the correction terms neglected will become important. Instead, we directly compute the integral 
\begin{equation}\label{}
  \Gamma_\text{ep} \sim \frac{\gamma }{k_F c_z \Lambda_z} \int_{q>0,\omega_z>0,\omega_z^2+C^2 q^{z_p-1} <T^2} \rd \omega_z  \rd q \frac{T}{\omega_D^2+\omega_z^2+C^2 q^{z_p-1}}\,.
\end{equation} Here the main contribution arises from thermally activated phonons, and $\omega_z=c_z q_z$. We perform a change of variable by $C^2 q^{z-1}=t^2$,  followed by $\omega_z=\rho \cos \theta$, $t=\rho \sin \theta$, to obtain
\begin{equation}\label{}
  \Gamma_\text{ep}\sim \frac{\gamma T}{k_Fc_z \Lambda_z} \frac{1}{C^{\frac{2}{z_p-1}}} \int_0^{T} \rho^{\frac{2}{z_p-1}}\rd \rho \int_0^{\pi/2} \rd \theta \frac{\left(\sin\theta\right)^{\frac{2}{z_p-1}-1}}{\omega_D^2+\rho^2}\,.
\end{equation} The angular integral is always finite and leads to an $\calO(1)$ factor. The remaining $\rho$-integral shows different behavior for $z_p<3,z_p=3$ or $z_p>3$. When $z_p<3$, the integral is IR finite, yielding 
\begin{equation}\label{}
  \Gamma_\text{ep}^{z_p<3}\sim \frac{\gamma}{k_F c_z \Lambda_z} \left(\frac{T}{C}\right)^{\frac{2}{z_p-1}}\,. 
\end{equation} When $z_p=3$, the integral is logarithmic, yielding 
\begin{equation}\label{}
  \Gamma_\text{ep}^{z_p=3} \sim \frac{\gamma}{k_F c_z \Lambda_z} \frac{T}{C}\ln\frac{T}{\omega_D}\,.
\end{equation} When $z_p>3$, the integral is IR divergent, yielding 
\begin{equation}\label{}
  \Gamma_\text{ep}^{z_p>3} \sim \frac{\gamma T}{k_F c_z \Lambda_z} \frac{\omega_D^{\frac{2}{z_p-1}-1}}{C^{\frac{2}{z_p-1}}}\,.
\end{equation}

The boundary of the regime is still given by Eq.\eqref{eq:B26}.

\subsubsection{Electron-Phonon critical regime (C)}

Here, the main contribution to $\Gamma_\text{ep}$ arises from the overdamped phonons that satisfy $\eta_\vec{q}\omega_\vec{q}^2\ll \gamma T$, and the gap term $\omega_D^2$ should be smaller compared to the dispersion term $C^2 q^{z_p-1}$, so we have 
\begin{equation}\label{}
  \Gamma_\text{ep} \sim \frac{\gamma }{k_F c_z \Lambda_z} \int_{0<q<\Lambda,0<\omega_z<c_z \Lambda_z,\omega_z^2+C^2 q^{z_p-1} <\frac{\gamma T}{v_F q}} \rd \omega_z  \rd q \frac{T}{\omega_D^2+\omega_z^2+C^2 q^{z_p-1}}\,.
\end{equation} In the range of $\omega_D,T$ of interest, we can assume $c_z\Lambda_z\gg \sqrt{\gamma T/(v_F q)}$, so we perform the $\omega_z$-integral first 
\begin{equation}\label{}
  \Gamma_\text{ep}\sim \frac{\gamma T}{k_F c_z \Lambda_z} \int_0^{q_1} \frac{\rd q}{\sqrt{\omega_D^2+C^2 q^{z_p-1}}} \arctan\sqrt{\frac{\frac{\gamma T}{v_F q}-C^2 q^{z_p-1}}{\omega_D^2+C^2 q^{z_p-1}}}\,,
\end{equation} where 
$$
q_1 = \left(\frac{\gamma T }{v_F C^2}\right)^{\frac{1}{z_p}}\,.
$$

When $z_p<3$, the integral is IR finite, we obtain 
\begin{equation}\label{}
  \Gamma_\text{ep}^{z_p<3}\sim \frac{\gamma T}{k_F c_z \Lambda_z} \frac{1}{C}q_1^{\frac{3-z_p}{2}} \sim \frac{\gamma T}{k_F c_z \Lambda_z} \frac{1}{C}\left(\frac{\gamma T}{v_F C^2}\right)^{\frac{3-z_p}{2z_p}}\,.
\end{equation} When $z_p=3$, the integral is logarithmic, yielding 
\begin{equation}\label{}
  \Gamma_\text{ep}^{z_p=3}\sim \frac{\gamma T}{k_F c_z \Lambda_z}\frac{1}{C} \ln\left(\frac{\gamma C T}{v_F \omega_D^3}\right)\,.
\end{equation}
When $z_p>3$, the integral is IR divergent, and needs to be cutoff at $q\sim q_\text{gap}\sim (\omega_D/C)^{2/(z_p-1)}$. The result is 
\begin{equation}\label{}
  \Gamma_\text{ep}^{z_p>3}\sim \frac{\gamma T}{k_F c_z \Lambda_z} \frac{1}{\omega_D} \left(\frac{\omega_D}{C}\right)^{\frac{2}{z_p-1}}\,.
\end{equation}

The crossover boundaries of the regime is the same as the pure 2D case. 

\subsubsection{Equipartition regime (D) and (E)}

In this regime, all phonons are thermally excited, and therefore 
\begin{equation}\label{}
  \Gamma_\text{ep} \sim {\frac{\gamma}{k_F c_z \Lambda_z}}\int_0^{c_z \Lambda_z}\rd \omega_z \int_0^{\Lambda}\rd q \frac{T}{\omega_D^2+\omega_z^2+C^2 q^{z_p-1}}\,.
\end{equation} The analysis of the integral is similar to previous subsections. We obtain for $z_p\leq 3$
\begin{equation}\label{}
  \Gamma_\text{ep}^{z_p<3} \sim \frac{\gamma T}{k_F} \frac{\Lambda^{\frac{3-z_p}{2}}}{C c_z \Lambda_z}\,,
\end{equation}
\begin{equation}\label{}
  \Gamma_\text{ep}^{z_p=3} \sim \frac{\gamma T}{k_F c_z \Lambda_z C} \ln \frac{C \Lambda}{\omega_D}\,.
\end{equation} When $z_p>3$, the equipartition is also enlarged and merges regimes (B,C,D), and $\Gamma_\text{ep}$ is 
\begin{equation}\label{}
  \Gamma_\text{ep}^{z_p>3}\sim \frac{\gamma T}{k_F c_z \Lambda_z} \frac{1}{\omega_D} \left(\frac{\omega_D}{C}\right)^{\frac{2}{z_p-1}}\,.
\end{equation}

\subsection{3D FS + 3D phonon}

Now we consider the case where the fermions and the phonons all disperse in 3D. The phonon dispersion we consider is 
\begin{equation}\label{}
  \omega_\vec{q}^2=C^2\vn{q}^{z_p-1}+\omega_D^2\,.
\end{equation}

We perform a similar manipulation on Eq.\eqref{eq:Gammaep_app}, by introducing an auxiliary variable $\vec{k'}=\vec{k}+\vec{q}$. The only difference is that the angular integral is replaced by the integral on the sphere:
\begin{equation}\label{}
  \int \sin\theta_q \rd \theta_q \rd \phi_q \sin \theta_{k'} \rd \theta_{k'} \rd \phi_{k'} \delta(\vec{k'}-\vec{q}-\vec{k})=\frac{2\pi}{\vn{k'}\vn{k}\vn{q}}\,.
\end{equation} As a result, we obtain 
\begin{equation}\label{eq:B60}
  \Gamma_\text{ep}=\frac{\pi\calN_\text{3D}\lambda^2}{k_F^2} \int_0^{\Lambda}\rd \vn{q} \vn{q} I(\beta,\omega_\vec{q},\eta_\vec{q})\,. 
\end{equation} The difference with the pure 2D case is the additional $\vn{q}$ factor that appears due to dimensionality. The different momentum crossovers described in Fig.~\ref{fig:I} still applies, and as a result, the crossover boundary will be the same as previous cases, and $\Gamma_\text{ep}$ can be estimated as the following.

\subsubsection{FL regime (A)}

\begin{equation}\label{}
  \Gamma_\text{ep} \sim \frac{\gamma}{k_F^2} \left[T \int_0^{q_1} \frac{q \rd q}{\omega_q^2}+\frac{\gamma T^2}{v_F}\int_{q_1}^{\Lambda} \frac{\rd q}{\omega_q^4}\right]\,,
\end{equation} where $q_1\sim \frac{\gamma T}{v_F \omega_D^2}\ll q_\text{gap}$. Here we have rewritten $\lambda$ using $\gamma=\pi \lambda^2\calN_\text{3D}$. The first term is proportional to $T^3$ and therefore subleading. The dominant contribution arises from the second term, which is 
\begin{equation}\label{}
  \Gamma_\text{ep} \sim \frac{\gamma^2 T^2}{k_F^2 v_F} \begin{cases}
                                                         \frac{1}{\omega_D^4} \left(\frac{\omega_D}{C}\right)^{\frac{2}{z_p-1}}, &  z_p>3/2\\
                                                         \frac{\Lambda^{3-2z_p}}{C^4}, & 1<z_p<3/2
                                                       \end{cases}
\end{equation}

\subsubsection{Generalized Bloch-Gr\"{u}neisen regime (B)}
Here, the contribution arises from thermally activated phonons in the regime $T\gg \max(\omega_D,\omega_*)$, and the typical phonon momentum is $q_T\sim (T/C)^{2/(z_p-1)}$.
\begin{equation}\label{}
  \Gamma_\text{ep} \sim \frac{\gamma T}{k_F^2} \int_0^{q_T} \frac{q \rd q}{\omega_D^2+C^2 q^{z_p-1}}\,. 
\end{equation}  The integral is IR finite when $z_p<3$, yielding 
\begin{equation}\label{}
  \Gamma_\text{ep}^{z_p<3} \sim \frac{\gamma}{k_F^2} \frac{T^{\frac{4}{z_p-1}-1}}{C^{\frac{4}{z_p-1}}}\,.
\end{equation} When $z_p=3$, the result is logarithmic 
\begin{equation}\label{}
  \Gamma_\text{ep}^{z_p=3}\sim \frac{\gamma T}{k_F^2 C^2} \ln\frac{T}{\omega_D}\,. 
\end{equation} When $z_p>3$, the result is IR divergent, leading to the equipartition result 
\begin{equation}\label{eq:B66}
  \Gamma_\text{ep}^{z_p>3}\sim \frac{\gamma T}{k_F^2}\frac{1}{\omega_D^2} \left(\frac{\omega_D}{C}\right)^{\frac{4}{z_p-1}}\,.
\end{equation}

\subsubsection{Electron-phonon critical regime (C)}

Here, the contribution arises from overdamped phonons, whose typical momentum is bounded by $q_1\sim (\gamma T/(v_F C^2))^{1/z_p}$.

\begin{equation}\label{}
  \Gamma_\text{ep} \sim \frac{\gamma T}{k_F^2} \int_0^{q_1} \frac{q\rd q}{\omega_D^2+C^2 q^{z_p-1}}\,.
\end{equation}

We obtain 
\begin{equation}\label{}
  \Gamma_\text{ep}^{z_p<3} \sim \frac{v_F}{k_F^2} \left(\frac{\gamma T}{v_F C^2}\right)^{\frac{3}{z_p}}\,,
\end{equation}
\begin{equation}\label{}
  \Gamma_{\text{ep}}^{z_p=3} \sim \frac{\gamma T}{k_F^2 C^2} \ln \left(\frac{\gamma C}{v_F \omega_D^2}\frac{T}{\omega_D}\right)\,,
\end{equation} and the $z_p>3$ case is still given by Eq.\eqref{eq:B66}.

\subsubsection{Equipartition regimes (D) and (E)}

\begin{equation}\label{}
  \Gamma_\text{ep} \sim \frac{\gamma T}{k_F^2} \int_0^{\Lambda} \frac{q \rd q}{\omega_D^2+C^2 q^{z_p-1}}\,. 
\end{equation}

We obtain 
\begin{equation}\label{}
  \Gamma_\text{ep}^{z_p<3} \sim \frac{\gamma T}{k_F^2 C^2} \Lambda^{3-z_p}\,,
\end{equation}
\begin{equation}\label{}
  \Gamma_\text{ep}^{z_p=3} \sim \frac{\gamma T}{k_F^2 C^2} \ln \frac{C \Lambda}{\omega_D}\,,
\end{equation} and the $z_p>3$ case is given by Eq.\eqref{eq:B66}.

\section{Frequency dependence of $\Gamma_\text{ep}$}\label{sec:Gamma_omega}

In this appendix, we estimate the frequency dependence of the electron-phonon scattering rate $\Gamma_\text{ep}$. Our goal is not to derive the full scaling function of $\Gamma_\text{ep}(\omega,T)$, but rather to determine the parametric dependence on $\omega$ in the different regimes identified in Sec.~\ref{sec:electronphonon}. We begin with the case of a 2D circular Fermi surface.

After analytically continuing Eq.~\eqref{eq:Sigma_ep1} to real frequency, we obtain
\begin{equation}\label{}
  \Gamma_\text{ep}(\omega,\vec{k})\equiv -2\Im \Sigma^R_\text{ep}(\omega,\vec{k})
  =-\lambda^2\int\frac{\rd^3\vec{q}}{(2\pi)^3}\int \frac{\rd z}{2\pi} \left[n_F(z+\omega)+n_B(z)\right]A_F(z+\omega,\vec{k}+\vec{q})A_X(z,\vec{q})\,.
\end{equation}
We can treat the momentum integral in the same way as in Appendix.~\ref{sec:funcI}. Since the fermion spectral function $A_F$ integrates to a constant near the Fermi surface, the nontrivial $\omega$-dependence enters only through the difference of Fermi functions. We therefore write
\begin{equation}\label{}
  \Gamma_\text{ep}(\omega,\vec{k})= \Gamma_\text{ep}(0,\vec{k})+\delta \Gamma_\text{ep}(\omega,\vec{k})\,,
\end{equation}
where the first term was analyzed in Appendix.~\ref{sec:funcI}, and the second is
\begin{equation}\label{}
  \delta \Gamma^{\text{2D}}_\text{ep}(\omega,\vec{k})= \frac{2\lambda^2\calN_\text{2D}}{k_F}\int_0^{\Lambda} \rd \vn{q}_\text{2D}\, \delta I(\beta,\omega,\omega_\vec{q},\eta_\vec{q})\,,
\end{equation}
with
\begin{equation}\label{eq:deltaI}
  \delta I \equiv -\int \frac{\rd z}{2\pi} \left[n_F(z+\omega)-n_F(z)\right] A_X(z,\vn{q})\,.
\end{equation}
We analyze this extra term in the limit of $|\omega|\gg T$. 
The factor $n_F(z+\omega)-n_F(z)$ restricts the integration variable $z$ to a window of width of order $|\omega|$. The problem is therefore controlled by whether this window resolves the phonon spectral peak. The phonon spectral function $A_X$ is peaked at $z\sim \omega_{\vec q}$, with width $\delta z\sim \gamma/\eta_{\vec q}$. Thus the same low-energy scales that controlled the temperature dependence in Appendix.~\ref{sec:funcI}, namely $T_\text{FL}$ and $\omega_*$, also separate the qualitative behaviors of $\delta\Gamma_\text{ep}$.

When $|\omega|\ll \omega_*$, the integration window in Eq.~\eqref{eq:deltaI} does not resolve the peak of $A_X$. In that case the phonon is sampled off shell, just as in the virtual-scattering regimes of Appendix.~\ref{sec:funcI}, and we obtain
\begin{equation}\label{}
  \delta \Gamma_\text{ep} \sim \frac{\gamma}{k_F} \int_0^{\Lambda} \rd \vn{q} \frac{\omega^2 \gamma\eta_{\vec{q}}}{\omega^2 \gamma^2+\eta_{\vec{q}}^2 \omega_\vec{q}^4}\,.
\end{equation}
The phonon dispersion may then be approximated by $\omega_\vec{q}^2\approx \omega_D^2$ when $|\omega|\ll T_\text{FL}$, and by $\omega_\vec{q}^2\approx C^2 \vn{q}^{z_p-1}$ when $T_\text{FL}\ll |\omega| \ll \omega_*$. This yields
\begin{equation}\label{}
  \delta \Gamma_\text{ep} \sim \frac{\gamma^2 \omega^2}{k_F v_F \omega_D^4}\ln\frac{T_\text{FL}}{|\omega|}\,,\qquad |\omega|\ll T_\text{FL}\,,
\end{equation}
and
\begin{equation}\label{}
  \delta \Gamma_\text{ep} \sim \frac{v_F}{k_F} \left(\frac{\gamma |\omega|}{v_F C^2}\right)^{\frac{2}{z_p}}\,,\qquad T_\text{FL}\ll|\omega|\ll \omega_*\,.
\end{equation}
These are parametrically the same as the $T$ dependences in regimes (A) and (C), upon replacing $T\to |\omega|$.

We move on to regime (B), In this opposite limit, $|\omega|\gg \omega_*$, the integration window in Eq.~\eqref{eq:deltaI} resolves the phonon spectral peak. The integral is then dominated by the peak of $A_X$, whose total weight scales as $\sim 1/\omega_{\vec q}$. This gives the estimate
\begin{equation}\label{eq:C8}
  \delta \Gamma_\text{ep} \sim \frac{\gamma}{k_F}\int_0^{\q_\omega}\rd \vn{q}\, \frac{1}{\omega_\vec{q}}\,.
\end{equation} Here, the upper limit $q_\omega$ is determined by the external frequency $\omega$, via the relation 
\begin{equation}
    q_\omega:~\omega_\vec{q}=\omega\,,
\end{equation} which yields $q_\omega\sim (\omega/C)^\frac{2}{z_p-1}$\,.  We obtain the estimate 
\begin{equation}\label{eq:C10}
    \delta \Gamma_\text{ep} \sim \frac{\gamma}{k_F}\frac{\omega^{\frac{3-z_p}{z_p-1}}}{C^{\frac{2}{z_p-1}}}\,,\qquad z_p<3\,.
\end{equation} The scaling exponent here also agrees with the scaling exponent of $\Gamma_\text{ep}$ in $T$. However, there is one difference, that the threshold for $z_p$ differs. In $\Gamma_\text{ep}$, the $T$-dependence turns logarithmic when $z_p=2$, and becomes dominated by IR when $z_p>2$. In contrast, in Eq.\eqref{eq:C8}, the threshold is $z_p=3$. More generally for spatial dimension $d$, the threshold is $z_c^\omega=2d-1$. When $z_p>z_c^\omega$, the integral is dominated by small $\vn{q}\sim (\omega_D/C)^\frac{2}{z_p-1}$, and we obtain 
\begin{equation}\label{eq:C11}
\delta\Gamma_\text{ep}\sim \frac{\gamma}{k_F}\frac{\omega_D^{\frac{3-z_p}{z_p-1}}}{C^{\frac{2}{z_p-1}}}\,,\qquad z_p>3\,,
\end{equation} and at $z_p=3$ we expect a logarithmic $\ln(\omega/\omega_D)$. 

Finally, we comment on the equipartition regimes (D) and (E). In regime (D), the $\omega$-dependence can still be estimated by Eq.\eqref{eq:C8}, by setting $q_\omega=\Lambda$ (momentum cutoff). As a result, $\delta\Gamma_\text{ep}$ becomes $\omega$-independent. However, now we expect $\delta \Gamma_\text{ep}\ll \Gamma_\text{ep}$, because the population of thermally excited phonons still grow with $T$, and when $T\gg \omega_\vec{q}$ this becomes the dominant contribution to the total $\Gamma_\text{ep}$.

Next, in regime (E), there is no simple comparison. On the one hand, the main contribution to $\Gamma_\text{ep}$ is the thermally excited phonons at the bottom of the band with an overall linear-in-$T$ dependence, and on the other hand, the contribution to $\delta \Gamma_\text{ep}$ is described by Eqs. \eqref{eq:C10} and \eqref{eq:C11}. Which one dominates depends on the specific values of $T$ vs $\omega$.

The  observation can be generalized to other scenarios with different dimensionalities, i.e. in regimes (A,B,C) the frequency dependence can be inferred from the $T$-dependence by replacing $T\to\omega$, but regime (B) has a different threshold $z_c^\omega=2d-1$; The frequency dependence saturates in regime (D) and beecomes much smaller than the linear-in-$T$ piece; And in regime (E) there is no simple comparison.

\section{Feedback from phonon to electronic criticality}\label{sec:feedback}

In this appendix, we estimate the leading feedback of the softened phonon on the electronic critical mode $\varphi$ within the Yukawa-SYK framework of Appendix.~\ref{sec:YSYK}. The quantity of interest is the $\varphi$ self-energy generated by the nonlinear coupling $u$,
\begin{equation}\label{}
  \Pi_{\varphi}^{(u)}(i\Omega,\vec{q})=u^2\int \frac{\rd \nu}{2\pi} \sum_{\vec{p}} D_\varphi(i\nu,\vec{p}) D_X(i\nu+i\Omega,\vec{p}+\vec{q})\,.
\end{equation}
We focus on the static part $\Pi_{\varphi}^{(u)}(i\Omega=0,\vec{q})$, which renormalizes the dispersion of the $\varphi$ boson. The question is whether this correction is subleading, comparable to, or more relevant than the bare $\varphi$ dispersion $\vn{q}^{z_{\varphi}-1}$.

To probe the most infrared-sensitive limit, we take the phonon to be fully softened and set $\omega_D=0$. This choice maximizes the low-energy phase space of the phonon and therefore provides the sharpest test of whether phonon feedback can destabilize the electronic critical dynamics.

We first consider scenarios (i) and (iii), in which the electronic and phonon sectors have the same dimensionality, $d_\text{el}=2$ or $3$. Substituting the propagators gives
\begin{equation}\label{}
  \Pi_\varphi^{(u)}(i\Omega,\vec{q})\sim \frac{u^2}{a_z^{3-d_\text{el}}}\int \rd^{d_\text{el}}\vec{p} \rd \nu \frac{1}{C^2\vn{p}^{z_p-1}+\frac{\gamma}{v_F}\frac{|\nu|}{\vn{p}}} \frac{1}{v_\varphi^2\vn{p+\vec q}^{z_\varphi-1}+\frac{\gamma_g}{v_F}\frac{|\nu|}{\vn{p+\vec q}}}\,.
\end{equation}
In the infrared, the dominant loop momenta satisfy $\vn{p}\sim \vn{q}$. The two propagators then introduce the frequency scales
\[
|\nu_1|\sim \frac{C^2 v_F}{\gamma}\,\vn{q}^{z_p}, \qquad
|\nu_2|\sim \frac{v_\varphi^2 v_F}{\gamma_g}\,\vn{q}^{z_\varphi}.
\]
For the cases of interest, $z_p<z_\varphi$, so $|\nu_1|\gg |\nu_2|$ at small $\vn{q}$. In that regime, the $v_\varphi^2\vn{p+\vec q}^{z_\varphi-1}$ term in the second denominator may be neglected over the dominant part of the frequency integral, which then becomes logarithmic between the scales $\nu_2$ and $\nu_1$. This gives
\begin{equation}\label{}
  \Pi_\varphi^{(u)}(i\Omega=0,\vec{q}) \sim \frac{u^2 v_F}{a_z C^2 \gamma_g}\,\vn{q}^{2+d_\text{el}-z_p} \ln \left(\frac{C^2 \gamma_g}{v_\varphi^2 \gamma}\vn{q}^{\,z_p-z_\varphi}\right)\,.
\end{equation}
Using Eq.~\eqref{eq:zp_res}, the power of the prefactor is
\[
2+d_\text{el}-z_p = z_\varphi-1,
\]
so this correction is marginal with respect to the bare $\varphi$ dispersion $\vn{q}^{z_\varphi-1}$. The logarithm is therefore the leading indication of how the softened phonon feeds back on the electronic critical mode. For $z_p<z_\varphi$, the logarithm grows in the infrared, which is consistent with an upward renormalization of the effective dynamical exponent $z_\varphi$.

Scenario (ii), with quasi-2D electrons and 3D phonons, is qualitatively different. In that case we find
\begin{equation}\label{}
  \Pi_{\varphi}^{(u)}(i\Omega=0,\vec{q}) \sim \vn{q_\text{2D}}^{\frac{7-z_p}{2}} \ln \vn{q_\text{2D}}\,,
\end{equation}
which is irrelevant compared to the bare $\varphi$ dispersion near $z_\varphi=3$ and $z_p=2$. Thus, in scenario (ii), the feedback of the softened phonon on the electronic critical mode is parametrically unimportant at low energies.

At leading order in $u^2$, the feedback is therefore marginal in scenarios (i) and (iii), but irrelevant in scenario (ii). This supports the main-text conclusion that the same cases which are most favorable for softened-phonon transport are also the most sensitive to feedback effects that can drive the electronic critical dynamics away from the marginal regime. A complete determination of the ultimate infrared flow would require a full renormalization-group treatment of the coupled electron-critical-mode-phonon system. It is nevertheless suggestive that Refs.~\cite{THolder2015a,THolder2015}, in a purely electronic theory, found the same qualitative trend toward a larger effective $z_\varphi$. Since the present problem involves an additional coupling to the phonon sector, however, it is not yet clear whether that trend should be identified directly with the feedback effects discussed here. Clarifying whether and how these mechanisms are related is an interesting problem for future work.

\section{Electron-phonon scattering rate in the presence of disorder}\label{app:disorder}

In this appendix, we justify the scaling forms for $\Gamma_\text{ep}$ quoted in Sec.~\ref{sec:disorder}. The point is that, once disorder becomes important, the low-energy regime relevant for phonon softening also modifies the phase-space counting in the electron-phonon scattering problem. We focus on the disorder-dominated regime
\[
v_\varphi\vn{q}\ll \Gamma,
\]
where the disorder-induced correction to the phonon self-energy discussed in Sec.~\ref{sec:disorder} is active. Assuming $v_\varphi\sim v_F$, this is also the regime in which the clean kinematic analysis of Appendix.~\ref{sec:funcI} must be modified.

To illustrate the argument explicitly, we consider the case of 2D electrons coupled to a 2D softened phonon. We start from the electron-phonon scattering rate in Eq.~\eqref{eq:Gammaep1},
\begin{equation}\label{eq:Gammaep_app}
    \Gamma_\text{ep}(\vec{k})=-\lambda^2 \int\frac{\rd^3 \vec{q}}{(2\pi)^3} \int \frac{\rd z}{2\pi} \frac{1}{\sinh \beta z} A_F(z,\vec{k}+\vec{q}) A_X(z,\vec{q})\,.
\end{equation}
Here $A_F$ and $A_X$ are the fermion and phonon spectral functions, obtained from
\begin{equation}\label{eq:Gfermion_disorder}
G(i\omega,\vec{k})^{-1}=i\omega+\frac{i\Gamma\sgn\omega}{2}-\xi_\vec{k}\,,
\end{equation}
and
\begin{equation}
    D_X^{-1}(i\Omega,\vec{q})=C'^2\vn{q}^{z_p'}+\frac{\gamma|\Omega|}{\Gamma}\,.
\end{equation}
In writing the fermion propagator, we retain only the elastic broadening from disorder. For the phonon propagator, we assume that in the low-energy regime of interest the dispersion is dominated by the disorder-renormalized self-energy contribution discussed below Eq.~\eqref{eq:DX_disorder}.

The key difference from the clean calculation is that the disorder broadening in Eq.~\eqref{eq:Gfermion_disorder} cuts off the kinematic enhancement associated with nearly on-shell scattering. In Appendix.~\ref{sec:funcI}, this enhancement produced the factor $2/(k_F\vn{q})$ appearing in Eq.~\eqref{eq:Gammaep_B3}. In the disorder-dominated regime, that factor is replaced by
\[
\frac{v_F}{k_F\Gamma},
\]
because the width of the fermion spectral function is now set by $\Gamma$ rather than by the clean momentum mismatch. Equivalently, the effective phase space for phonon scattering is reduced by a factor of order $v_F\vn{q}/\Gamma$ compared to the clean case.

In the overdamped phonon regime, this gives
\begin{equation}
    \Gamma_\text{ep}\sim \frac{\gamma v_F}{k_F} T \int_0^{q_1'}\frac{\vn{q}}{\Gamma}\frac{\rd q}{\omega_q^2}\,.
\end{equation}
The extra factor $\vn{q}/\Gamma$ makes explicit the disorder-induced suppression of the scattering phase space. The phonon dispersion entering this expression is
\[
\omega_q^2=C'^2 q^{z_p'},
\]
and the upper limit $q_1'$ is fixed by the boundary of the overdamped regime,
\begin{equation}
    C'^2(q_1')^{z_p'}\sim \frac{\gamma T}{\Gamma}\,.
\end{equation}
Substituting this scale into the momentum integral, we obtain
\begin{equation}
    \Gamma_\text{ep}\sim \frac{v_F}{k_F}\left(\frac{\gamma T}{\Gamma C'^2}\right)^{\frac{2}{z_p'}}\,,\qquad \text{when }z_p'<2\,,
\end{equation}
and
\begin{equation}
    \Gamma_\text{ep} \sim \frac{\gamma}{k_F } \frac{\Gamma T}{\omega_D^2 v_F}\,,\qquad\text{when }z_p'>2.
\end{equation}
Thus, in the disorder-dominated regime $T<T_\Gamma$ the low-temperature scattering rate is weaker than in the clean theory.

It is useful to express the result in terms of the clean exponent $z_p$. Using Eqs.~\eqref{eq:zp_res} and \eqref{eq:zpp}, we find
\begin{equation}\label{eq:zpp2}
    z_p'=z_p-2\,.
\end{equation}
Disorder therefore lowers the effective exponent controlling the softened phonon dynamics in this regime, which is why the temperature dependence of $\Gamma_\text{ep}$ becomes less singular than in the disorder-free case.

The same analysis can be generalized to 3D electrons coupled to a 3D phonon. The extra momentum-space dimension contributes one additional power of $q$ to the phase-space integral, so the low-temperature scattering rate scales as
\begin{equation}
    \Gamma_\text{ep}\propto T^{\frac{3}{z_p'}}\,,
\end{equation}
while the relation between the disorder-controlled and clean exponents remains
\[
z_p'=z_p-2.
\]
Thus, in both 2D and 3D, disorder introduces an additional low-energy suppression of electron-phonon scattering, even though it can simultaneously enhance the tendency toward phonon softening discussed in Sec.~\ref{sec:disorder}.

\newpage

\end{appendix}





\bibliography{refs}


\end{document}